\documentclass[prb,twocolumn,superscriptaddress]{revtex4}
\usepackage{amsfonts}
\usepackage{amsmath}
\usepackage{amssymb}
\usepackage{graphicx}
\usepackage{tabularx}
\usepackage{subfigure}
\DeclareMathOperator{\erf}{erf}

\newcommand{\beq}{\begin{equation}} 
\newcommand{\eeq}{\end{equation}}
\newcommand{\bea}{\begin{eqnarray}}
\newcommand{\eea}{\end{eqnarray}}
\begin{document}

\title{Temperature-Dependent Behavior of Confined Many-electron Systems in the Hartree-Fock Approximation}
\author{Travis Sjostrom} 
\affiliation{Quantum Theory Project, Dept.\ of Physics and Dept.\ of %
Chemistry, University of Florida, Gainesville FL 32611}
\author{Frank E.\ Harris}
\affiliation{Quantum Theory Project, Dept.\ of Physics and Dept.\ of %
Chemistry, University of Florida, Gainesville FL 32611}
\affiliation{Dept. of Physics, University of Utah, Salt Lake City UT 84112}
\author{ S.B.\ Trickey}
\affiliation{Quantum Theory Project, Dept.\ of Physics and Dept.\ of %
Chemistry, University of Florida, Gainesville FL 32611}

\begin{abstract}
Many-electron systems confined at substantial finite temperatures and
densities present a major challenge to density functional theory.
Comparatively little is known about the free-energy behavior of such
systems for temperatures and pressures of interest, for example, in
the study of warm dense matter.  As a result, development of
approximate free-energy functionals is faced with difficulties of
assessment and calibration.  Here we address, in part, this need for
detailed results on well-characterized systems.  We present results on
a comparatively simple, well-defined, but computationally feasible
model, namely the thermal Hartree-Fock approximation applied to eight
one-electron atoms with nuclei at fixed, arbitrary positions in a
hard-walled box. We discuss the main technical tasks (defining a
suitable basis and evaluation of the required matrix elements) and
discuss the physics which emerges from the calculations.
\end{abstract}

\date{4 October 2011}

\maketitle

\section{Introduction}

\label{sec1}

Warm dense matter (WDM) is encountered in systems as diverse
as the interiors of giant planets \cite{Fortney08,MilitzerHubbard09} 
and in the pathway to inertial confinement fusion \cite{Lindl95,McGroryEtAl08}. 
WDM is challenging to theory and simulation 
because it occurs inconveniently, for theory, between the 
comparatively well-studied plasma 
and condensed matter regimes.  Both the 
Coulomb coupling parameter $\Gamma := Q^2/(r_s k_B {\rm T}) $ and
electron degeneracy parameter $\Theta := k_B {\rm T}/\varepsilon_F$
are approximately unity for WDM.  ($Q =$ relevant charge, $r_s =$
Wigner radius, $\varepsilon_F =$ electron Fermi energy, ${\rm T} =$
temperature, $k_B =$ Boltzmann constant.)  A non-perturbative
treatment therefore is required.  

Contemporary computations on WDM 
\cite{Alavi94,Silvestrelli99,Surh01,Desjarlais02,Galli04,Mazevet04,%
Mazevet05,Recoules06,Faussurier09,Horner09,Recoules09,Vinko09,Wunsch09,%
Clerouin10} are dominated by use of the 
Kohn-Sham (KS) realization of thermal density functional theory (DFT)
\cite{Mermin65,GuptaRajagopal82,PPLB82,Perdew85,StoitsovPetkov88,%
Dreizler89,Eschrig10} to generate a potential surface for 
ionic motion (treated classically).  The great majority of these 
calculations use 
approximate ground-state exchange-correlation (XC) functionals, $E_{xc}$, 
with the temperature dependence of the XC free energy picked up 
implicitly from the
$\rm T$-dependence of the density $n({\mathbf r},{\rm T})$.  
Though fruitful, this approach is not without potential difficulties,
as is illustrated in Fig.\ 3 of Ref.\ \onlinecite{Surh01}.  Three 
issues are germane here.  

First, virtually nothing systematic is known about the implicit 
$\rm T$-dependence of ground-state approximate $E_{xc}$ functionals 
(especially beyond the local density approximation, LDA), 
whether they be constraint-based or empirical.  Compared to the
ground-state situation, there is only a small
literature on explicitly $\rm T$-dependent functionals, that is, 
XC free energy functionals, and essentially all of those studies are 
at the level of the LDA  
\cite{Perrot79,Gupta80,PerrotDw84,Tanaka85,Kanhere86,Dandrea86,Stolzmann87,%
Yonei87,Cler92,Geldart93,Perrot94,Cler95,PerrotDw00,Dw03,Lambert06,%
Lambert07,Ritchie07}.

Second, there is the computational burden of solving for the Kohn-Sham
orbitals and eigenvalues. Since the computational load from the
eigenvalue problem scales, in general, as order $N_{\rm orbitals}^3$,
the growth in the number of non-negligibly occupied KS orbitals with
increasing temperature is a clear computational bottleneck.  See the
remarks, for example, in Sec.\ 4 of Ref.\ \onlinecite{Clerouin10}.
For complicated systems, the same bottleneck is encountered in
ground-state simulations which use the DFT Born-Oppenheimer energy
surface to drive the ionic dynamics.  One result has been the
emergence of active research on orbital-free DFT (OFDFT), that is,
approximate functionals for the ingredients of the KS free energy,
namely the KS kinetic energy (KE) ${\mathcal T}_s$, entropy 
${\mathcal  S}_s$, and XC free energy $\mathcal{F}_{xc}$ or their 
ground-state counterparts.  Almost all of this effort has been for 
ground-state OFKE functionals \cite{PRB80,GHDS10,HuangCarter10,HungEtAl10}.  
(Note that most of the OFKE literature invokes the KS separation of the KE
in order to use existing $E_{xc}$ approximations consistently.)

Third, the finite-temperature OFDFT work is dominated by variants on 
Thomas-Fermi-von Weizs\"acker theory; see for example Ref.\ 
\onlinecite{Lambert07} and references therein.  That type of 
theory, however, is known (on both fundamental and computational
grounds) to be no more than qualitatively accurate in
many circumstances relevant to WDM ({\it e.g.}, chemical
binding). Compared to the data-rich context for development of zero-temperature
functionals, there is little to guide development and assessment of finite-T
functionals. Similarly, compared to the T=0 situation
(or the very high T situation), not much is known about the accuracy 
of approximate finite-T OFDFT functionals beyond TFvW. An aim of the present
study is to provide such data for the combination of a well-defined 
physical system with a well-defined approximation and its implementation. 

\section{System and Methodology}

\label{sec2}

We shall treat a neutral system of $N_{ion}$ atoms with 
nuclei fixed
at arbitrary positions in a hard-walled three-dimensional
rectangular box. The confined 
system allows systematic treatment of pressure effects at stipulated
finite temperature, hence is a small, treatable sample of WDM.  
For specificity in the initial work, we chose one-electron atoms
and $N_{ion} \le 8$ for simplicity.  Note the
corresponding use of a small number of atoms in Ref.\ \onlinecite{Clerouin10}.
Also note the considerable literature on spherically
confined systems at ${\rm T}=0$ K 
\cite{LeSarHerschbach81,Pang94,CruzSoullardGamaly99,BielinskaEtAl01,%
CruzSoullard04,AdvQC57}.  
We have not found any work on  
lower-symmetry confinement of 
multi-atom systems at non-zero $\rm T$.  At non-zero $\rm T$ there is 
a large literature on average-atom methods, for example Refs.\ 
\onlinecite{Fromy96,Clerouin10} and many others, but such methods 
do not provide the fiduciary data needed for functional development. 

As a many-electron problem, a clearly defined 
approximation is required.  We choose 
the finite temperature Hartree-Fock (FTHF)
scheme \cite{StoitsovPetkov88,Mermin63,Sokoloff67,DolbeaultEtAl09}, 
with issues of electron correlation to be addressed in the
future. FTHF provides a clear definition for
the correlation free energy which is consistent with
the definition used in zero-temperature theory with
approximate wave functions.  (Of course, the definition of
correlation energy in zero-temperature DFT differs, as it
does in finite-temperature DFT, but that
is no more a barrier here than there.) In addition to being 
well-defined in the grand ensemble, FTHF provides the
advantage that its ${\rm T}=0$ K limit is the {\it lingua franca} 
of molecular electronic structure interpretation.  Use of 
FTHF therefore also provides at 
least a semi-quantitative framework for understanding 
chemical processes in WDM.  

The FTHF approximation is defined in the grand canonical ensemble 
by restricting the relevant traces to states which are single
Slater determinants \cite{StoitsovPetkov88,Mermin63,Sokoloff67}.  
The result is an upper bound to the free energy 
$\mathcal{F}_{FTHF} \ge \mathcal{F}$. 
Standard thermodynamic relationships for the grand ensemble follow.   
The FTHF Euler equations to be solved (in unrestricted form) are
\cite{Sokoloff67}  
\begin{align}
\varepsilon_i\varphi_i(\mathbf{r}) = &\left( -\frac{1}{2}\nabla^2 %
+ U_{ion}(\mathbf{r}) \right) \varphi_i(\mathbf{r}) \nonumber \\%
&+  \sum_j{f_j\int{d\mathbf{r^{\prime}} %
\frac{\left|\varphi_j(\mathbf{r^{\prime}})\right|^2}%
{\left|\mathbf{r}-\mathbf{r^{\prime}}\right|}}\varphi_i(\mathbf{r})} \nonumber \\
&- \sum_j {\delta_{\sigma_i\sigma_j} f_j \int{d\mathbf{r^{\prime}} %
\frac{\varphi_j^{*}(\mathbf{r^{\prime}})\varphi_i(\mathbf{r^{\prime}})}%
{\left|\mathbf{r}-\mathbf{r^{\prime}}\right|}}\varphi_j(\mathbf{r})} \, , 
\label{HFeigvaleqn}
\end{align}
with $U_{ion}$ the ion-electron interaction potential and $\sigma_i$
the spin label; the sums are over all spin orbitals.
Unless indicated otherwise we 
use hartree atomic units ($\hbar = m_e = e =1$; energy is then
in hartrees,
l hartree = 27.2116 eV,  and lengths are in bohrs, 1 bohr = 0.52918 angstrom).
The spin orbitals (eigenstates, $\varphi_i$) have Fermi-Dirac thermal occupations 
\begin{equation}
f_i = \left( 1+e^{\beta(\varepsilon_i-\mu)} \right)^{-1} \;  , \quad %
 N=\sum_i f_i\;,
\label{FDocc}
\end{equation}
where $\beta = 1/k_B{\rm T}$ and $\mu$ is the electron chemical potential. 
For a specified  value of $N$, which is a grand ensemble average, $\mu$
must be determined.   These equations, along with specification of the 
nuclear sites and imposition of hard-wall boundary conditions, completely
describe the problem. 

The FTHF free energy and entropy are given by
\begin{equation}
  \mathcal{F}_{FTHF} = \sum_i{f_i\varepsilon_i}-
\frac{1}{2}\sum_{i,j}f_if_j\left( J_{i,j}-K_{i,j} \right) 
-{\rm T}{\mathcal S_{FTHF}} \;,
\label{FreeEnDefn}
\end{equation}
\begin{equation}
  \mathcal{S}_{FTHF} = -k_B \sum_i{f_i\ln(f_i)+(1-f_i)\ln(1-f_i)}\;,
\label{EntropyDefn}
\end{equation}
with conventional definitions of $J$ and $K$: 
\begin{align}
J_{i,j}:= &  \int d\mathbf{r} d\mathbf{r^{\prime}} %
\frac{ \left| \varphi_i(\mathbf{r})\right|^2 %
\left|\varphi_j(\mathbf{r^{\prime}})\right|^2} %
{\left|\mathbf{r}-\mathbf{r^{\prime}}\right|}\;, \\
K_{i,j} := & \delta_{\sigma_i\sigma_j}
\int  d\mathbf{r} d\mathbf{r^{\prime}} %
\frac{\varphi_i(\mathbf{r})\varphi_j(\mathbf{r})%
\varphi_i^{*}(\mathbf{r^{\prime}})\varphi_j^{*}(\mathbf{r^{\prime}})}%
{\left|\mathbf{r}-\mathbf{r^{\prime}}\right|}\;.
\label{JKdefns}
\end{align}
Equations (\ref{FreeEnDefn}) and (\ref{EntropyDefn}) clearly reduce to
conventional ground-state Hartree-Fock expressions in the zero-temperature
limit.   
%We reiterate that, though these expressions appear to be
%the ground-state Hartree-Fock equations with Fermi occupancies
%inserted, such is not the case.  

\subsection{Gaussian Basis Sets}

Solution of self-consistent field equations such as (\ref{HFeigvaleqn})
via Gaussian-type-orbital (GTO) basis methods is the standard 
procedure in modern 
computational codes for molecular systems. First introduced by Boys 
\cite{Boys50,Boys60,Singer60,LongstaffSinger60,Gill94}, such basis
sets automatically satisfy the free-molecule boundary 
condition that the orbitals vanish
at infinity.  For a hard-wall confined system, the basis functions must 
vanish at the boundary, so standard molecular GTO matrix-element expressions 
are inapplicable.
This fact illustrates the most critical technical issue for implementation, 
namely, to find a basis that satisfies the boundary conditions 
yet allows for an efficient enough evaluation of the two-electron 
integrals to be computationally tractable on reasonable resources.
A second, closely related technical issue is that the high temperature
also dictates what is ``efficient enough'', in that the basis must be 
large enough and flexible enough to represent a sufficient number of thermally
occupied higher-energy orbitals of the system, hence to represent the
density and free energy accurately.

Those considerations eliminate several seemingly plausible options 
for a basis. For
example, a real-space finite-difference/element scheme, while suitable
for a DFT calculation or a Hartree-Fock calculation on a free 
diatomic molecule (for which curvilinear coordinates can be
exploited \cite{KobusEtAl96}), is far too expensive for the present
case because of the  number of matrix
elements to be calculated. Another example is sine functions.  They 
also satisfy the
hard-wall boundary condition, but an adequate description of the rapidly
varying electronic distribution near the nuclei requires
prohibitively many matrix elements in our multi-center problems. 
So we chose an adaptation of standard GTO methods which uses 
modified Gaussians that meet the boundary condition, yet  retain
enough efficiency to complete the calculation. 

The various ways to force a GTO to zero at
the bounding planes of a rectangular box can have great 
impact upon the efficiency of the matrix element calculations. 
Compared with familiar practice for free molecules, in general 
the confined case requires more primitive GTOs for each contracted
one.  More importantly, the finite integration volume makes it
impossible to achieve  completely analytic calculation
of the two-electron integrals, which is, of course, precisely
the category in which computational efficiency
is most needed.
We have addressed this issue by using truncated Gaussians
as described in the next section.

\subsection{Truncated Gaussians}

The rectangular box makes Cartesian GTOs a convenient choice,  because
each primitive function then is separable into Cartesian factors which
are simple 1D functions. Consider the Cartesian factor 
\begin{equation}
  g^n(x)=(x-x_c)^n e^{-\alpha(x-x_c)^2}
\label{xstype}
\end{equation}
To force this function to zero at the box 
boundaries $x=0$ and $x=L_x$, we  subtract
a constant equal to the function value at each end. When $x_c$ is not
at the box center, the value to be subtracted differs for the two ends, 
so we split the function into two pieces, make the two subtractions, and 
scale the two pieces such that the resulting 
function is continuous. Each unnormalized Cartesian factor becomes
\begin{align}
  g^n_{box}(x) &= a_0 \left( g^n(x) - \Delta_0 \right)\quad 0\le x \le x_c %
\nonumber \\
  &=  a_L \left( g^n(x) - \Delta_{L_x} \right) \quad x_c\le x \le L_x 
\label{sgtotruncated}
\end{align}
with $\Delta_0 = g^n(0)$, $\Delta_{L_x} = g^n(L_x)$.
We call this the truncated GTO (tGTO) basis.

Two technical issues remain.  The tGTO functions
may not have continuous derivatives, so proper 
evaluation of the kinetic energy matrix elements requires attention. 
Appendix A shows that nothing untoward happens and that 
the kinetic energy is simply a sum of piecewise contributions, 
except for $p$-type functions which have a simple correction term.
Second is the matter of evaluating two-electron matrix elements.  In
Appendix B, we show that this task reduces to computing 
finite-range integrals of products of Gaussians and error functions.
At this juncture, we are doing those via Gauss-Legendre quadrature.
We also note that the tGTO basis is simpler than the smoothly cut-off
floating spherical GTO basis of Ref.\ \onlinecite{CruzSoullard04} in the 
sense that their cut-off introduces 
a mixing of two symmetry types ({\it e.g.} s, d) in each basis function.

Note that, so far, we have implemented only the 
restricted Hartree-Fock approximation (RHF; non-spin-polarized in
DFT language; closed-shell or spin-compensated in quantum chemistry language).

\section{Results}

\subsection{Zero Temperature} 

Two simple ground-state test cases, the H atom and
the H$_2$ molecule, illustrate the system behavior with
increasing confinement (decreasing volume) as well as the
correctness of our implementation. The confined-system
energies should be above the ground-state energies (in the 
basis selected) of the corresponding free systems and approach those 
free-system energies (and bond length for the molecule) in
the limit of large box volume.   

\begin{figure}
    \subfigure[]{\includegraphics[angle=-90,width=2.6in]{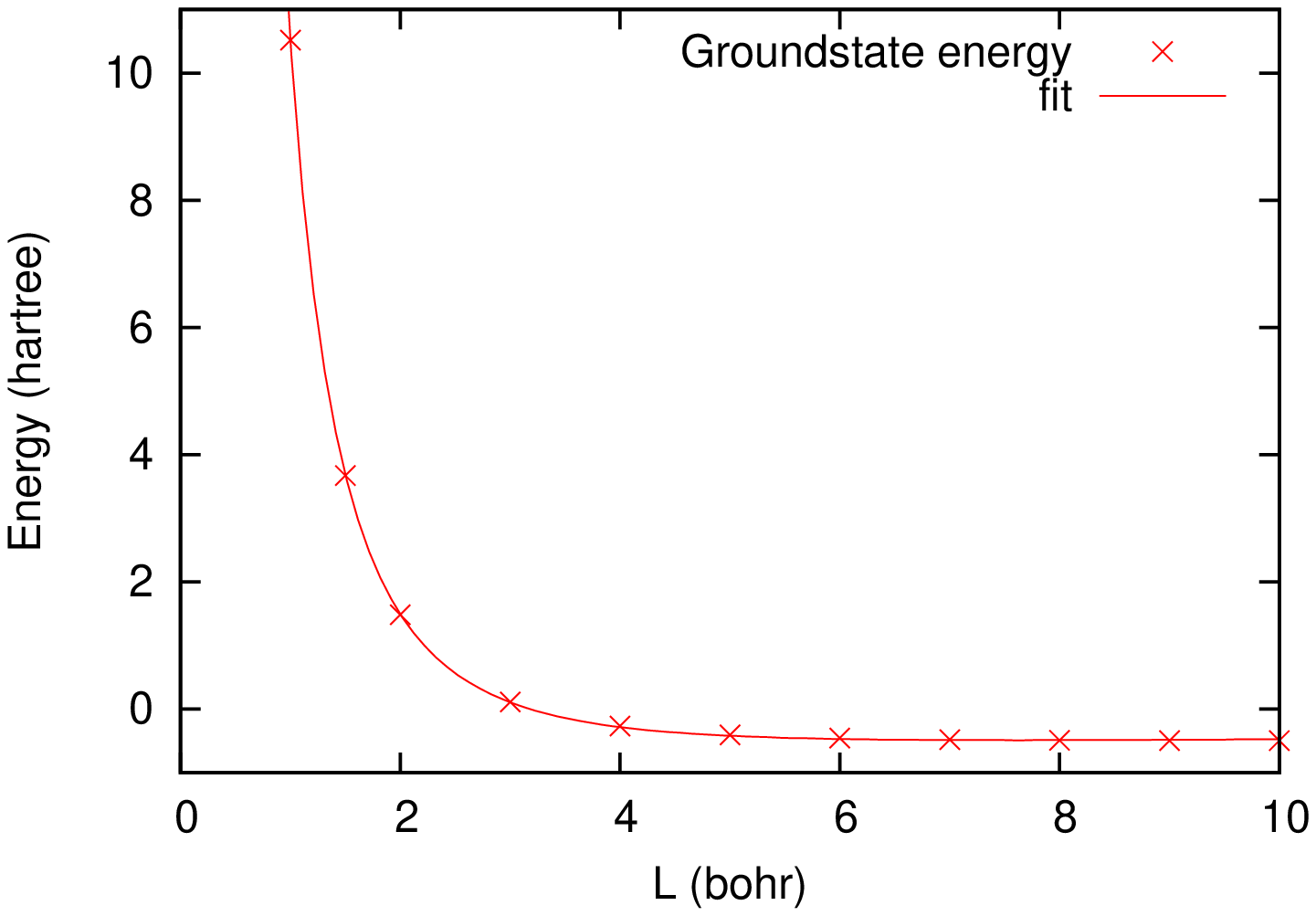}}
    \subfigure[]{\includegraphics[angle=-90,width=2.6in]{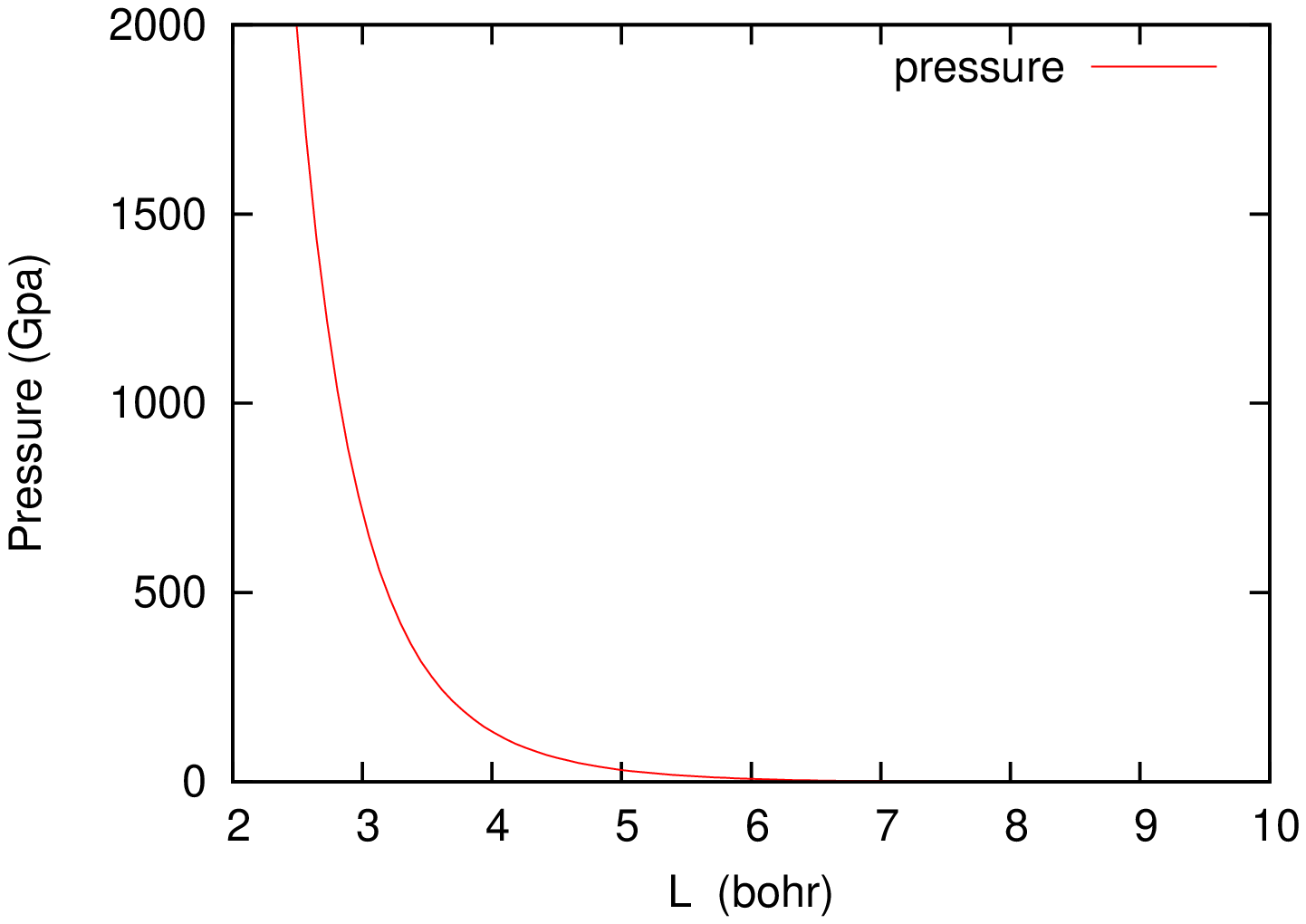}}
    \caption{(a) Ground-state energy of an H atom in the center of a cube with %
edge length $L$. Fit is for function $f(L) = a/L^2 + b/L + c$. %
(b) Pressure calculated from the energy fit.}    \label{hsize} 
  \end{figure}

For the hydrogen atom in the center of a cubical box, Fig.~\ref{hsize}
 (a) shows the ground-state energy for
confinements $L \le 10$ bohr. 
(The basis exponents shown in Table \ref{tbl1}; their selection
is described in the next section.)   
The calculations were carried out to $L=30$ bohr.  Beyond $10$ bohr there was 
negligible difference in the energy with respect to the free-atom
energy in the same basis, precisely as expected. At $L=30$ bohr the 
ground-state energy is identical with the free-atom GTO
calculation using the same basis, namely $-0.498476$
hartree, validating our overlap, kinetic, and nuclear energy integral
calculations.  A fit to the energies as a function of cube edge $L$
\beq
E(L) = a/L^2 + b/L + c  \; \; ,\quad 0 \le L \le 10 \,\, \mathrm{bohr} 
\eeq
yields $a= 14.5733$ hartree bohr$^2$, $b= -3.82369$, hartree bohr, $c= -0.238258$ hartree. 
Though this fit does not have the correct infinite-size limit, it is the 
best fit of this simple form for $0 \le L \le 10$ bohr.
>From this fit the pressure, $p = -dE / dV$ ($V=L^3$) can be calculated; see 
Fig.\ \ref{hsize} (b).

\begin{figure}
  \subfigure[]{\includegraphics[angle=-90,width=2.6in]{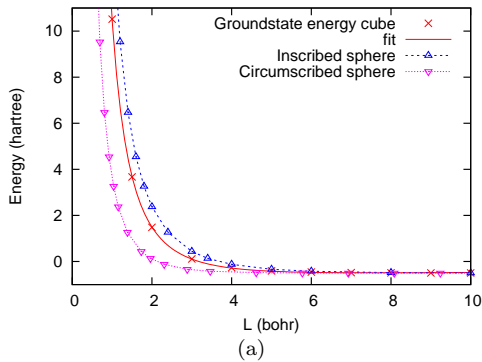}}
  \caption{Comparison of ground-state energy of an H atom in a cube
 with its energy within two bounding spheres.}
  \label{cube-sphere}
\end{figure}

\begin{figure}
  \subfigure[]{\includegraphics[angle=-90,width=2.6in]{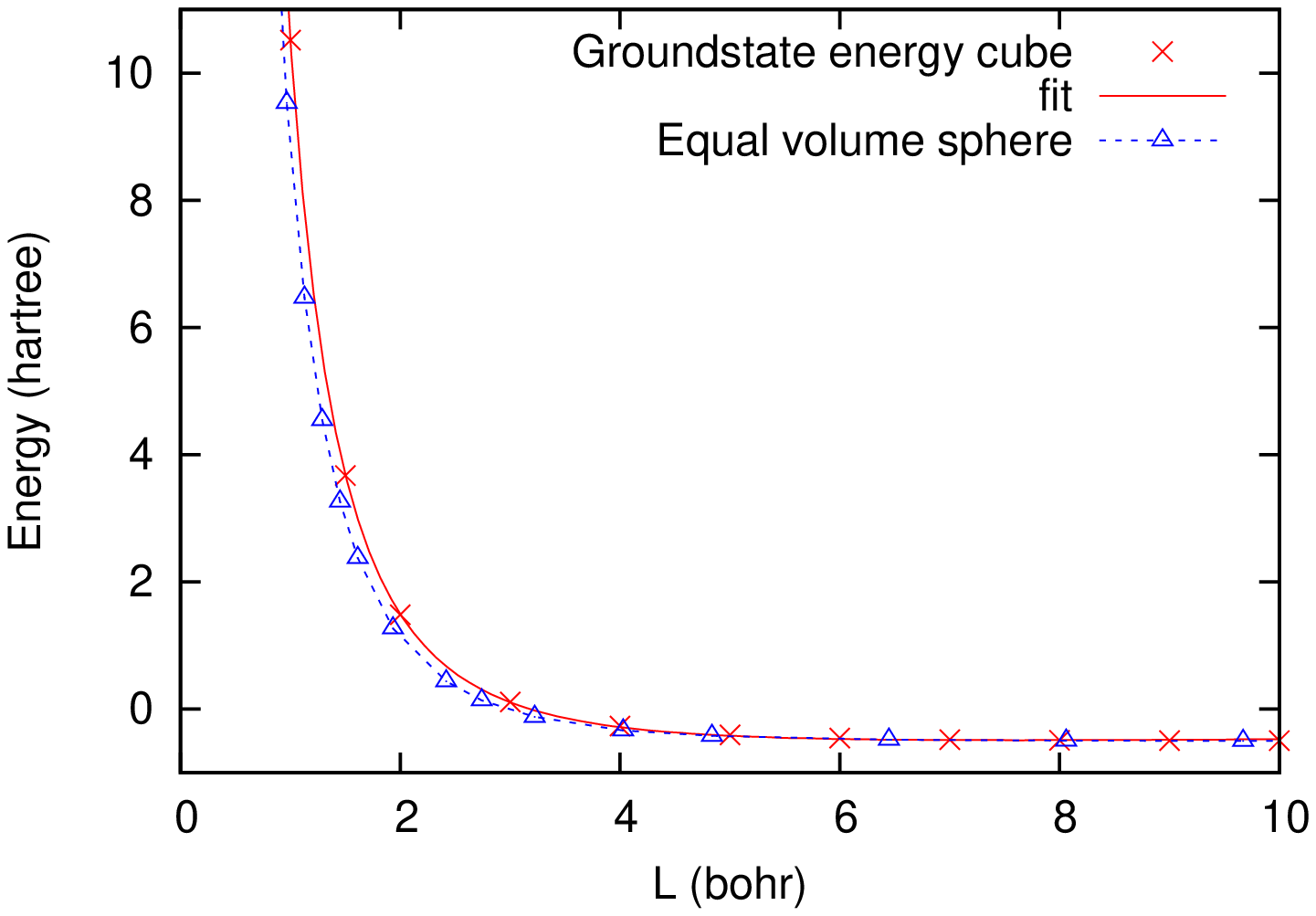}}
  \subfigure[]{\includegraphics[angle=-90,width=2.6in]{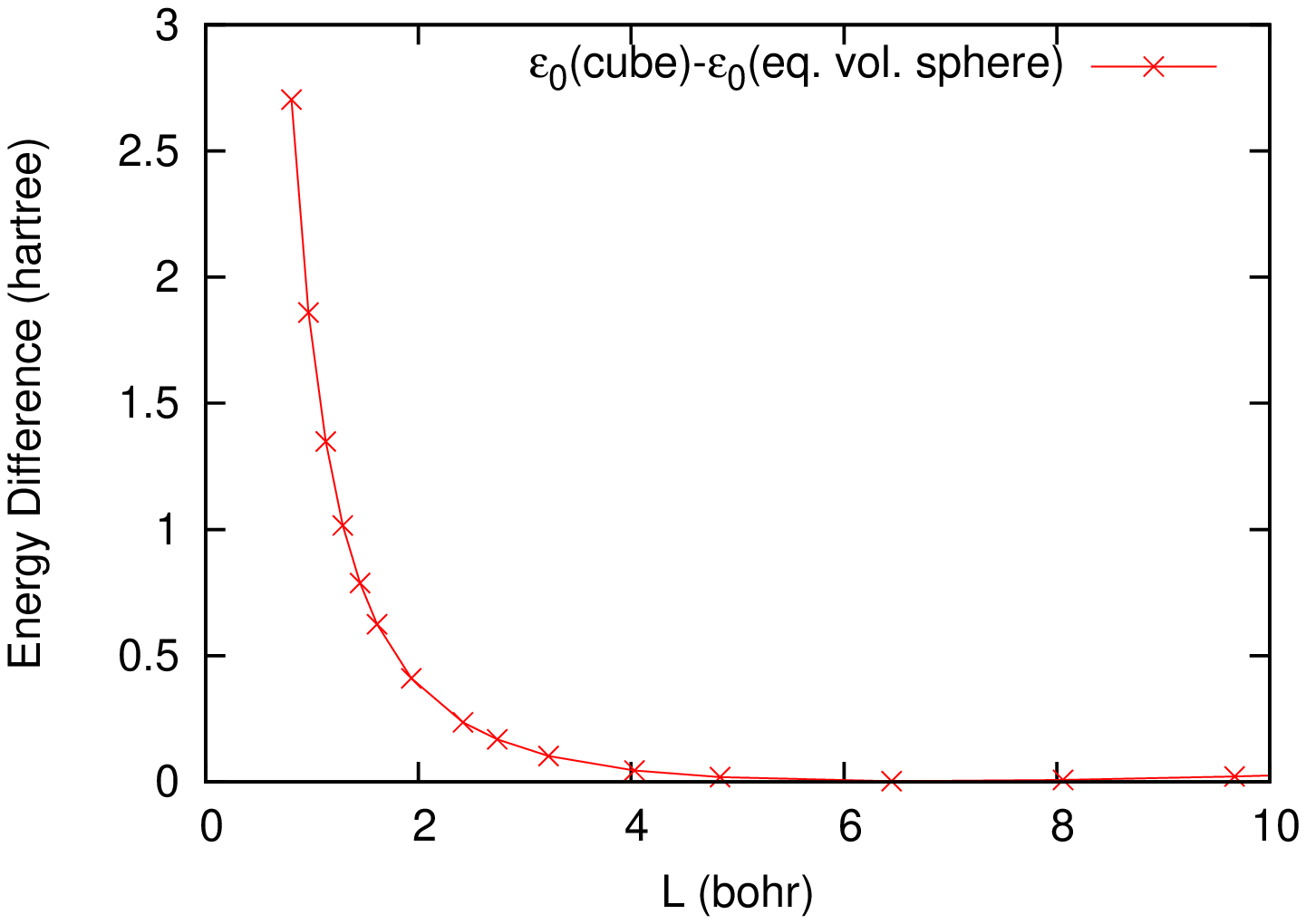}}
  \caption{(a) Comparison of energy of an H atom in a cube with that in
 a sphere of equal volume. (b) The difference of the
 two ground-state energies.}
  \label{cube-sphere2}
\end{figure}

The effects of spherical versus cubic confinement can be assessed easily by
comparison of the two confinement
types both for bounding volumes and equal volumes.  
For bounding volumes, the circumscribed-sphere $R= \sqrt{3}L/2 $ and
 inscribed-sphere
$R=L/2$  results from Ref. 
\onlinecite{AquinoEtAl07} (also see Ref. \onlinecite{Ludena78}) 
can be compared to our cube results.  Figure \ref{cube-sphere} 
shows that the energy of the cubically confined system is bounded by that of 
the two spherical systems, as 
expected.  Shape effects of confinement are shown in 
Fig.\ \ref{cube-sphere2}. Though the spherical system  
has a lower energy than that of the equal volume cubical system, 
they are quite close until the cube is smaller than $L\approx 4$ bohr, where
a significant
 indication of shape dependence begins. At $L=1$ bohr the difference
 in energy is over $25\%$, as shown in Fig. \ref{cube-sphere2} (b).
 A simpler basis of six $s$-type tGTO with exponents
[0.15,0.3,0.6,1.2,2.4,4.8 bohr$^{-2}$] reproduces the
 ground-state energies for $L \ge 1.5$ bohr.

\begin{figure}
    \includegraphics[angle=-90,width=2.6in]{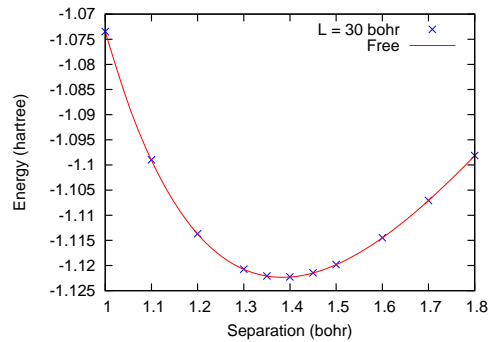}
    \caption{Energy versus atomic separation for the H$_2$ molecule %
 centered in a 30 bohr cube. The ground-state agrees exactly with the free GTO %
calculation using the same primitives as a basis.}
    \label{h2bond}
  \end{figure}

\begin{figure}
    \includegraphics[angle=-90,width=2.6in]{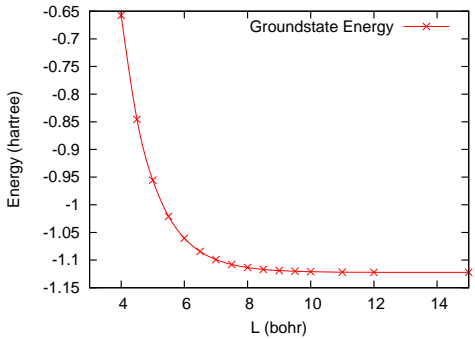}
    \caption{Ground-state energy for H$_2$ with bond length $1.4$ bohr 
confined in cube of edge length $L$.}
    \label{h2boxsize}
  \end{figure}

\begin{figure}
    \includegraphics[angle=-90,width=2.6in]{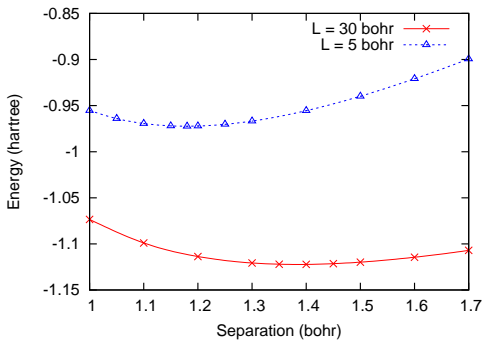}
    \caption{Energy vs.\ bond length for the H$_2$ molecule 
centered in cubes $L=30$ and $L = 5$ bohr. The minima are 
at $1.383$ and $1.178$ bohr respectively.}
    \label{h2bond2}
  \end{figure}

Next consider the confined H$_2$ molecule at zero temperature. Here we 
used the simpler six-tGTO basis just given.
For a large cube, the new tGTO 
confined-box computations again should conform to known 
results for the integrals and produce essentially the energy vs.\ bond length
 curve for the free molecule.  
With $L = 30$ bohr and the molecule 
centered in the cube and aligned along the body diagonal, we get
energies shown as points in Fig.\ \ref{h2bond}.  These 
agree completely with the values of the free GTO calculation, which
are shown in that figure as the ``Free'' curve. 
The minimum is at 1.383 bohr, satisfactorily 
close to the free-molecule RHF value of 1.385 bohr from a 
6-31G$^{**}$ basis calculation \cite{SzaboOstlund}.
Conversely for fixed $\mathrm R$ (at 1.4 bohr) and decreasing $L$ (to 4 bohr),  
the ground-state energy behavior is shown in Fig.\ \ref{h2boxsize}.
The onset of confinement effects becomes visible in the vicinity of 
$L \approx 11$ bohr.  Optimization of the bond length $\mathrm R$
 at $L = 5$ 
bohr is shown in Fig.\ \ref{h2bond2}.  The total energy is, of course,
higher than for the larger box, and the optimal $\mathrm R$ is 
shifted down to $1.178$ bohr from  $1.383$ bohr.

\begin{figure}
    \includegraphics[angle=-90,width=2.6in]{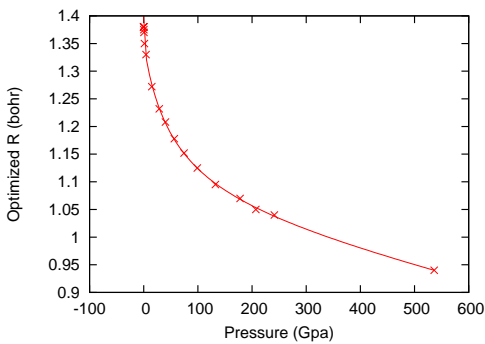}
    \caption{Optimized bond length, $\mathrm R$, for the H$_2$ molecule %
 vs.\ pressure.}
    \label{h2bl-p}
  \end{figure}

Following this method, we obtain the optimized $\mathrm R$
 and energy for decreasing $L$. A function fit to the energy
 similar to that used earlier yields
  $\mathrm R$ as a function of the
 pressure, as shown in Fig.\ \ref{h2bl-p}.

\subsection{Finite Temperature}

For finite-temperature calculations, at least a single {\it p} orbital
needs to be included, and as many orbitals as feasible should be
available to represent the fractionally occupied levels which 
become increasingly important as $\rm T$ is increased.  For all the 
following calculations, 
except where noted, the basis consists of seven {\it s}-type primitive
GTOs and one $p_x$, one $p_y$, and one $p_z$ GTO, with the $p$-GTO exponents
equal.  An elementary exponent optimization was done as follows.
A set of five $s$-type exponents was picked and held fixed:  
[0.1,0.2,0.4,0.8,1.6 bohr$^{-2}$]. The sixth exponent was optimized to 
to minimize $\varepsilon_{1s}$ for the single atom centered in a cube
of specified $L$. With those six exponents fixed, the seventh
$s$-GTO exponent was used to minimize $\varepsilon_{2s}$.  With those
seven fixed, the $p$ exponent was used to minimize $\varepsilon_{2p}$.
Additionally, for small $L$ the smallest exponents produce orbitals
that are so similarly flat that an approximate linear dependence exists and diagonalization fails. 
When this happens the smallest
exponent is replaced by extending the even-tempered
exponent series to larger values.
For example at $L=4$ the $0.1$ exponent
is replaced with $3.2$, at $L=2$ the $0.2$ exponent is also
replaced with $6.4$. 

This procedure keeps the ratio of the effective
length ($1/\sqrt{\alpha}$) of the most diffuse function to the edge 
length $L$ at 0.75-0.79 for $L \le 3$, with smaller ratios for larger
$L$.    

The optimization was done for each $L$. Table \ref{tbl1} shows the
resulting exponents and energies for the orbitals that were optimized.

\begin{table*}
  \begin{ruledtabular}
  \begin{tabular}{cccccccc}

    L & fixed & 1$s$ & 2$s$ & 2$p$ & $\varepsilon_{1s}$ & $\varepsilon_{2s}$ & $\varepsilon_{2p}$ \\ \hline

    1.0 & 1.6, 3.2, 6.4, 12.8, 25.6 & 179.1 & 244 (NA) & 4.01 & 10.518 & 50.5898 & 26.683 \\ 
    1.5 & 0.8, 1.6, 3.2, 6.4, 12.8 & 92.15 & 244 (NA) & 1.88 & 3.6738 & 21.3545 & 11.154  \\ 
    2 & 0.4, 0.8, 1.6, 3.4, 6.8 & 48 & 250 & 1.115 & 1.48471 & 11.3649 & 5.8753  \\ 
    3 & 0.2, 0.4, 0.8, 1.6, 3.4 & 24.5 & 175 & 0.545 & 0.11385 & 4.47073 & 2.25356 \\ 
    4 & 0.2, 0.4, 0.8, 1.6, 3.4 & 21.2 & 7.0 & 0.34 & $-$0.268848 & 2.18313 & 1.06314 \\ 
    5 & 0.1, 0.2, 0.4, 0.8, 1.6 & 11.2 & 3.7 & 0.235 & $-$0.40474 & 1.18372 & 0.547955 \\ 
    6 & '' & 10.4 & 2.5 & 0.18  & $-$0.458898 & 0.675591 & 0.288794 \\ 
    7 & '' & 17.7 & 2.25 & 0.15  & $-$0.481704 & 0.389716 & 0.145105 \\ 
    8 & '' & 10.2 & 0.195 & 0.122  & $-$0.491112 & 0.217062 & 0.0591604 \\ 
    9 & '' & 10.2 & 0.07 & 0.103  & $-$0.495291 & 0.10651 & 0.00450443 \\ 
    10 & '' & 10.2 & 0.0365 & 0.088  & $-$0.497104 & 0.0327616 & $-$0.0321549 \\ 
    11 & '' & 10.1 & 0.023 & 0.076  & $-$0.497889 & $-$0.0179536 & $-$0.0578267 \\ 
    12 & '' & 10.1 & 0.018 & 0.0665  & $-$0.498227 & $-$0.0535872 & $-$0.0764093 \\ 
    13 & '' & 10.1 & 0.015 & 0.059  & $-$0.498372 & $-$0.0789901 & $-$0.0901715 \\ 
    14 & '' & 10.05 & 0.014 & 0.053  & $-$0.498435 & $-$0.0972631 & $-$0.100500 \\ 
    15 & '' & 10.1 & 0.014 & 0.048  & $-$0.498461 & $-$0.108291 & $-$0.108291 \\
  \end{tabular}
  \end{ruledtabular}
  \caption{Optimized exponents and orbital energies for an H atom %
 at the center of a cube of edge length $L$. 2$p$ refers to the %
triply degenerate $p_x$, $p_y$, and $p_z$ states. For $L < 2$, %
 the 2$s$ level could not be optimized beyond the first 6 exponents. %
   \label{tbl1}}
\end{table*}

\begin{figure}[t]
  \includegraphics[angle=-90,width=2.6in]{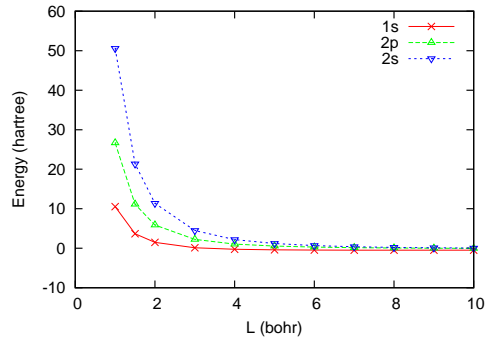}
   \caption{Lowest five eigenstate energies of H as a function of cube edge 
$L$. Energy values are from Table \ref{tbl1}.}
  \label{loweststates}
\end{figure}

\begin{figure*}
  \subfigure[]{\includegraphics[angle=-90,width=2.6in]{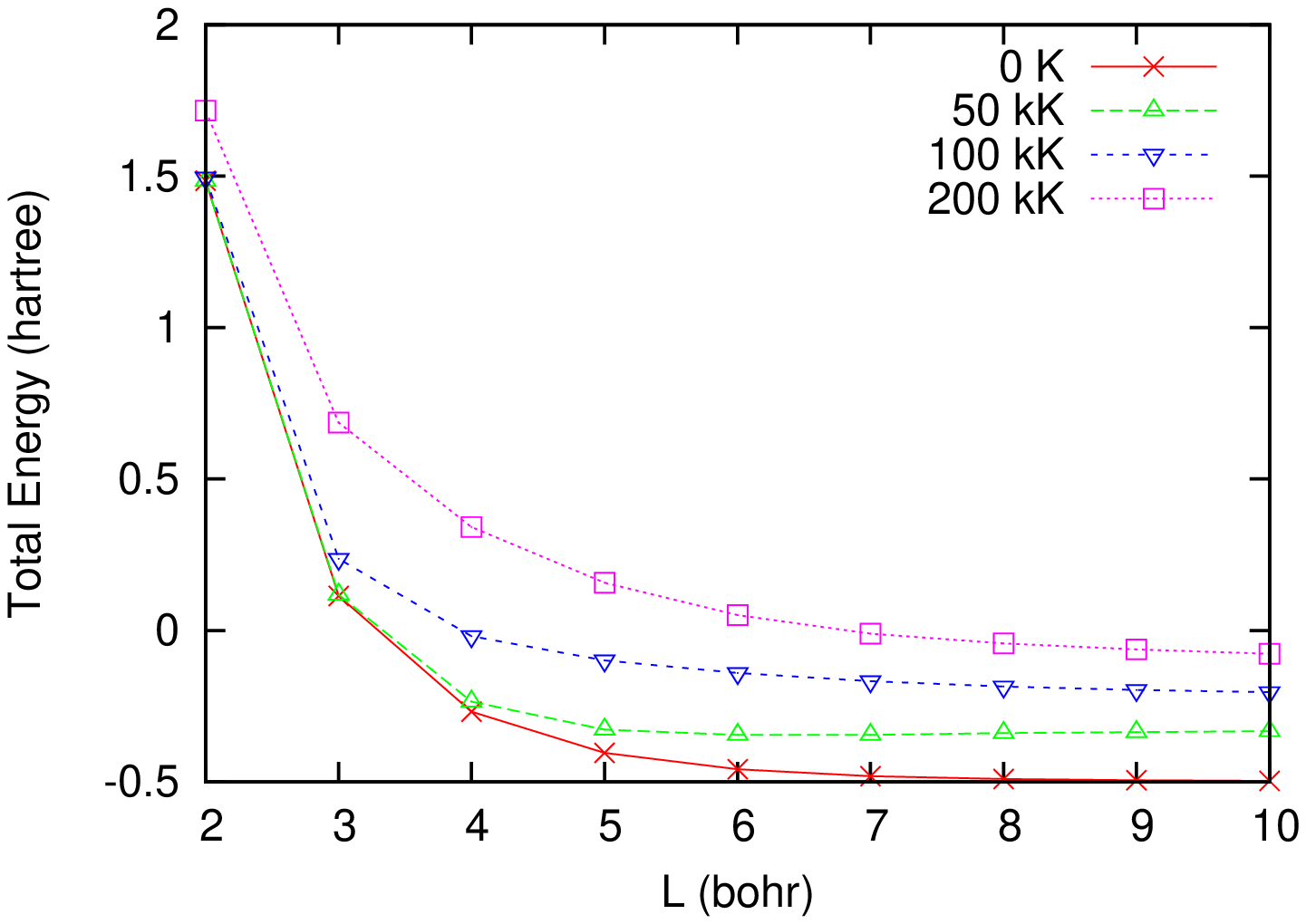}}
   \subfigure[]{\includegraphics[angle=-90,width=2.6in]{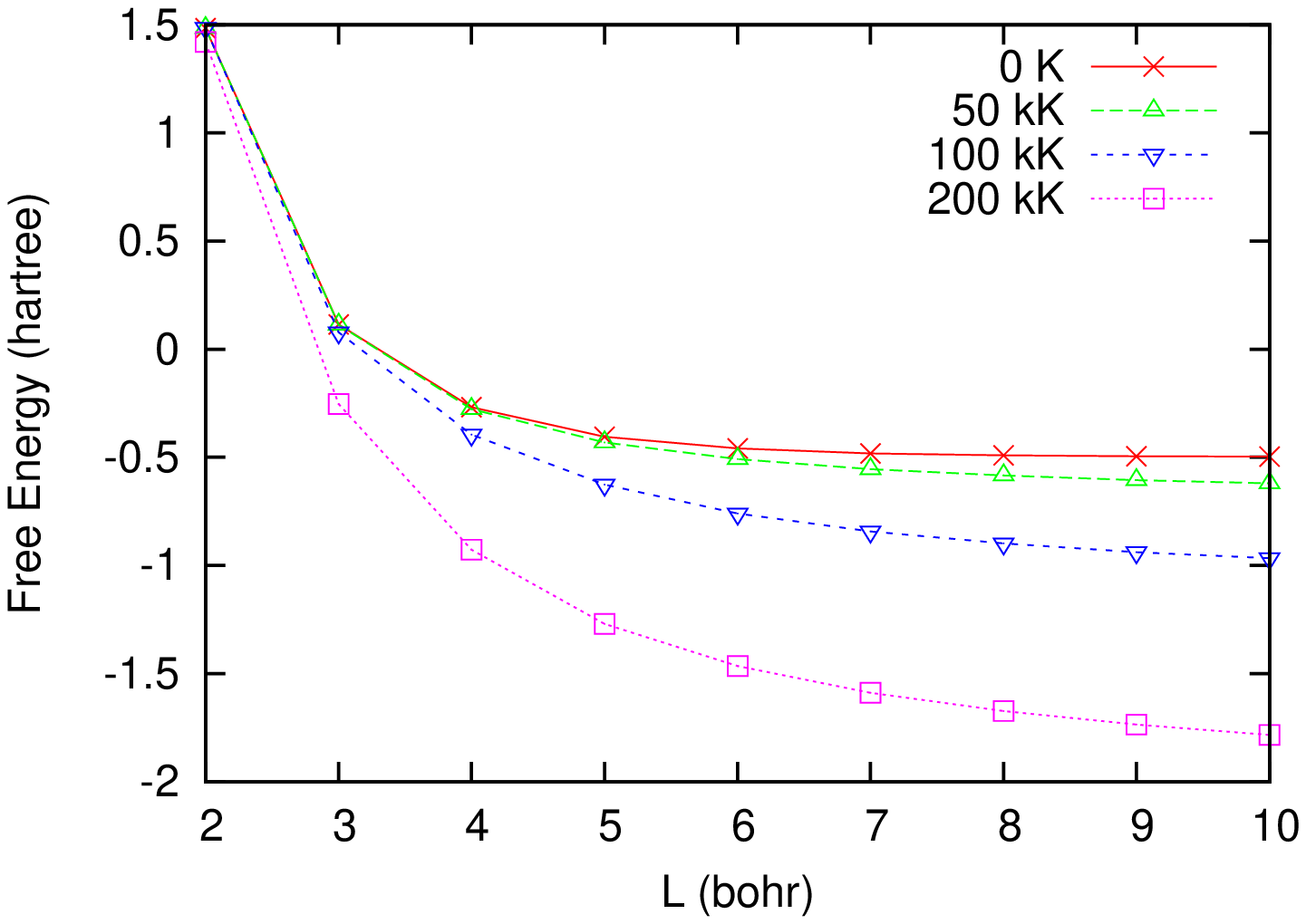}}
   \caption{(a) Single H atom total energy, %
$E={\mathcal F}_{FTHF} + {\rm T}{\mathcal S}_{FTHF}$, and (b) free energy, %
$\mathcal{F}_{FTHF}$ as a function of cube edge $L$, %
 with the one-electron levels populated according to the Fermi distribution.}
  \label{henergysize}
\end{figure*}

With this volume-dependent optimized basis,  calculations were
done for a single
atom at the center of a cube  with $1 \le L \le 15$ bohr.  The 
orbital energies of the
five lowest states (1$s$, 2$p_x$, 2$p_y$, 2$p_z$, 2$s$) are plotted in
 Fig.\ \ref{loweststates}.
Notice the inversion of ordering ($2p$ below $2s$) that is a result
primarily of the confinement. 
To address finite temperature for this single-electron system, the
one-electron levels were populated according to the Fermi-Dirac
distribution. Observe that the one-electron Hamiltonian is independent
of density, so the one-electron orbitals and eigen-energies are 
independent of occupancy, even though the density and total electronic
energy are not. The left-hand panel in Fig.\ \ref{henergysize} shows 
the resulting total energy as a function of $L$ for four values of 
$\rm T$, while the right-hand panel shows the free energy.
The weak minimum in total energy in 
the vicinity of  $6$ bohr at T=50 kK appears to be
a confinement effect.
  We have found a similar minimum
at about the same volume by doing a Fermi-Dirac population of the 
high-precision eigenvalues of the spherically confined H atom given 
in Ref.\ \onlinecite{AquinoEtAl07}.  

\begin{figure*}
  \subfigure[]{\includegraphics[angle=-90,width=2.6in]{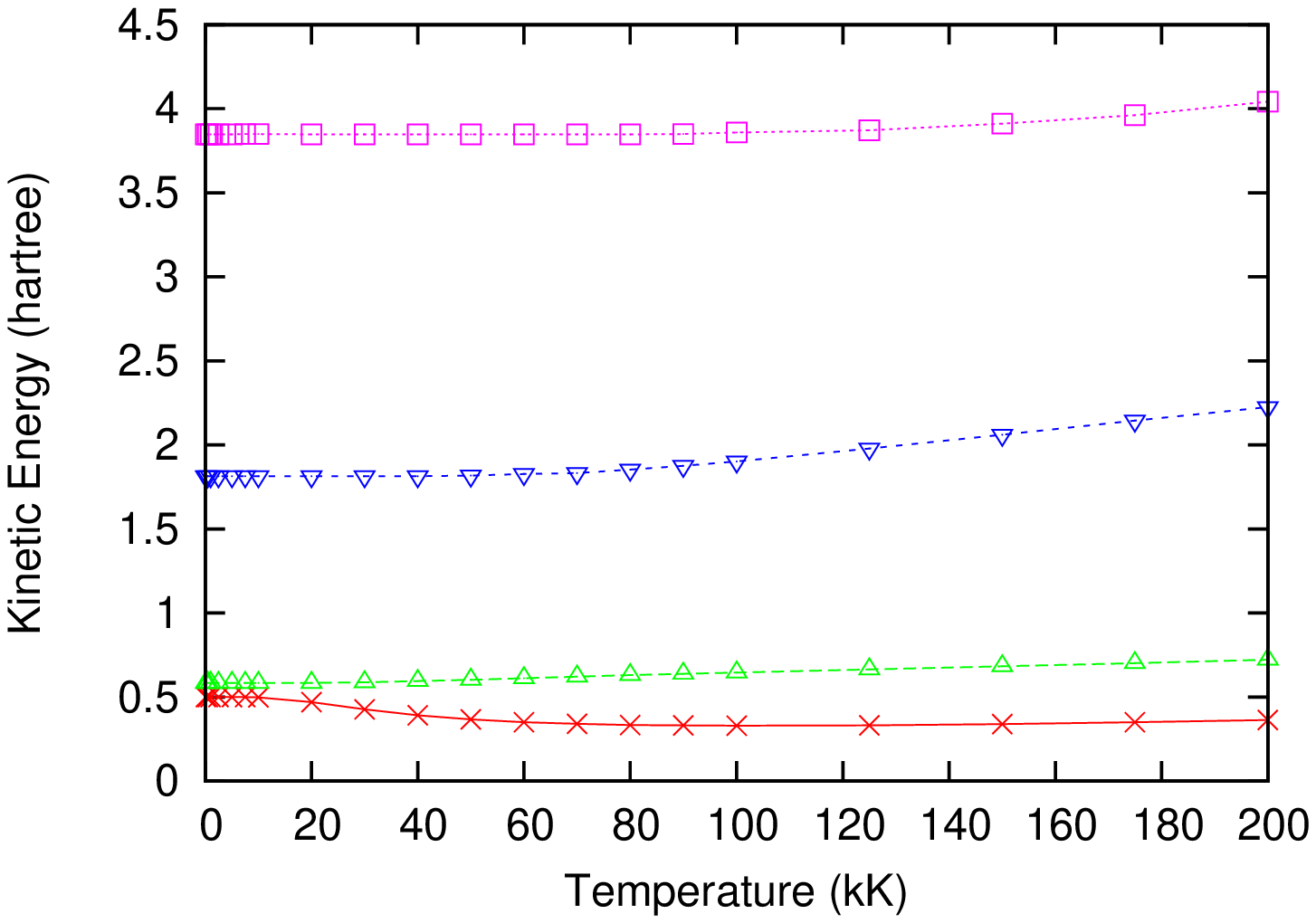}}
  \subfigure[]{\includegraphics[angle=-90,width=2.6in]{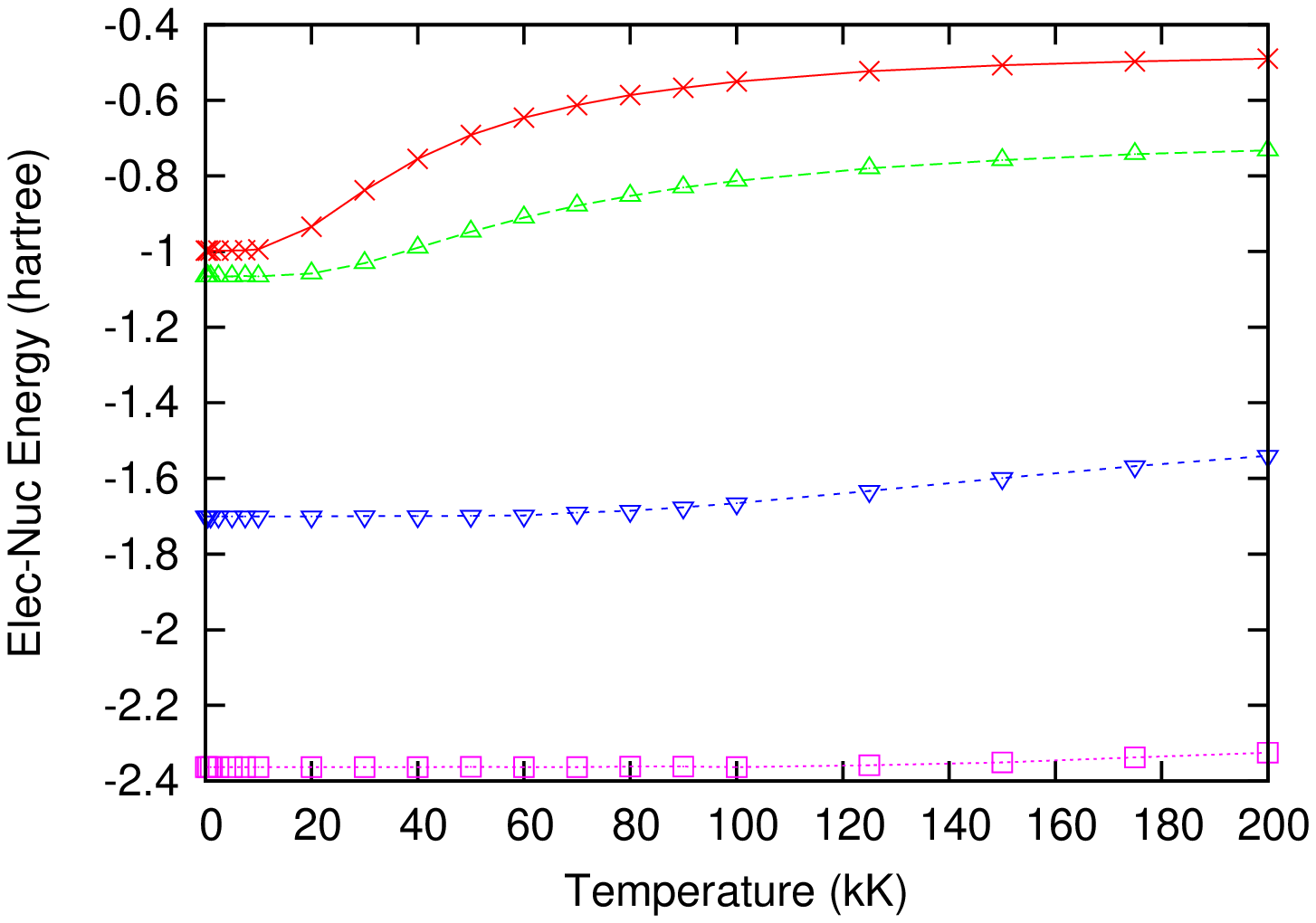}}
  \subfigure[]{\includegraphics[angle=-90,width=2.6in]{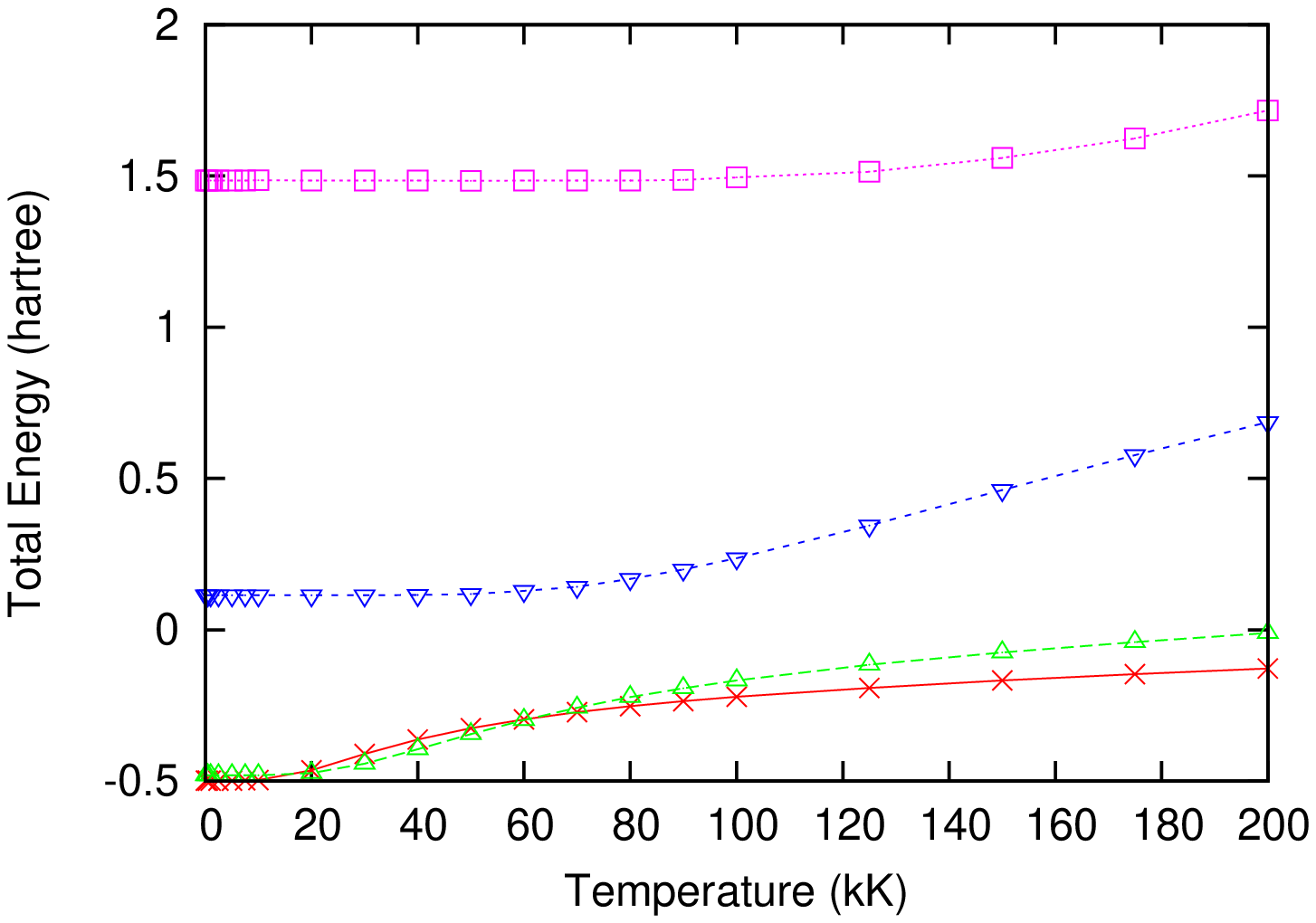}}
  \subfigure[]{\includegraphics[angle=-90,width=2.6in]{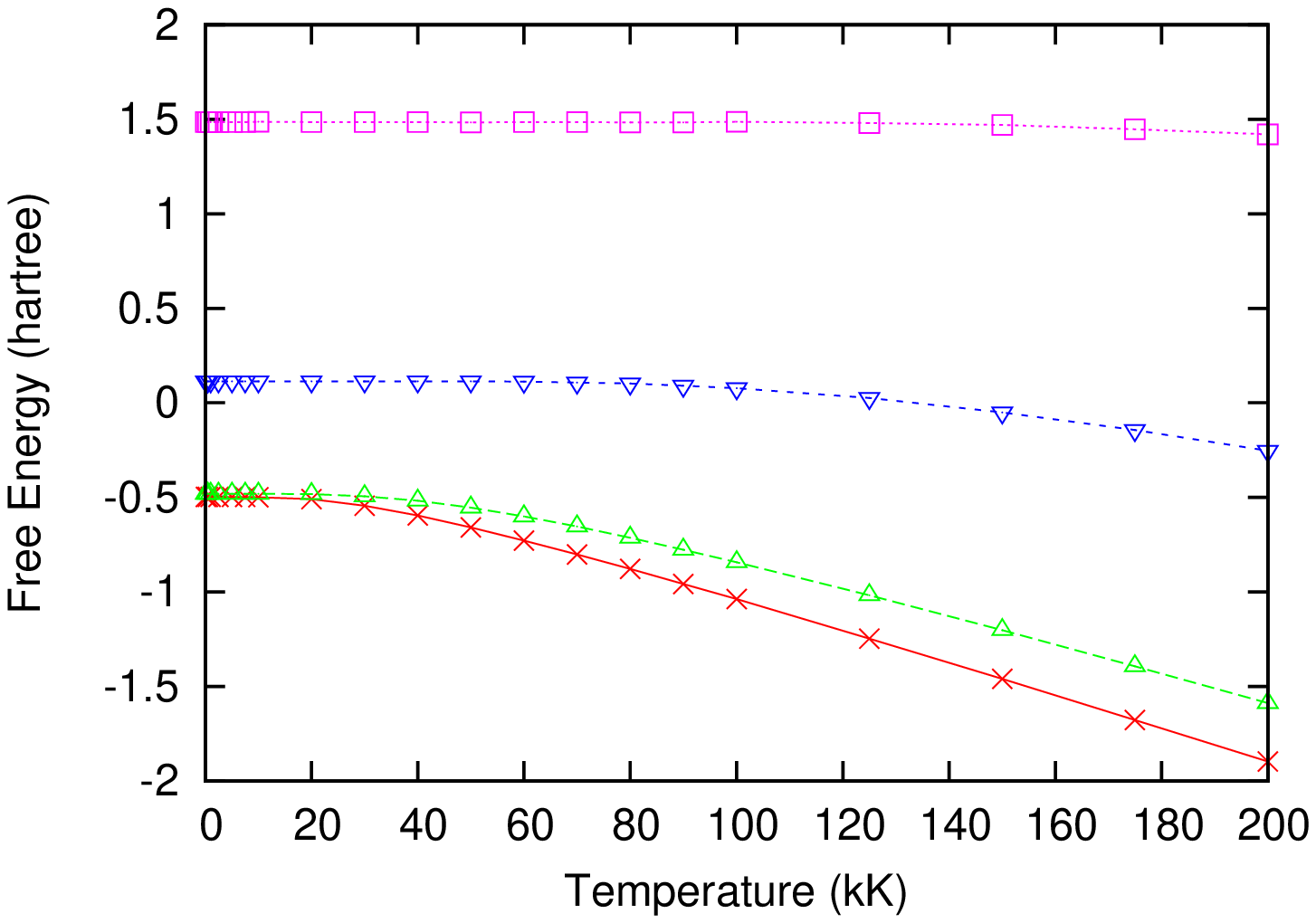}}
  \subfigure[]{\includegraphics[angle=-90,width=2.6in]{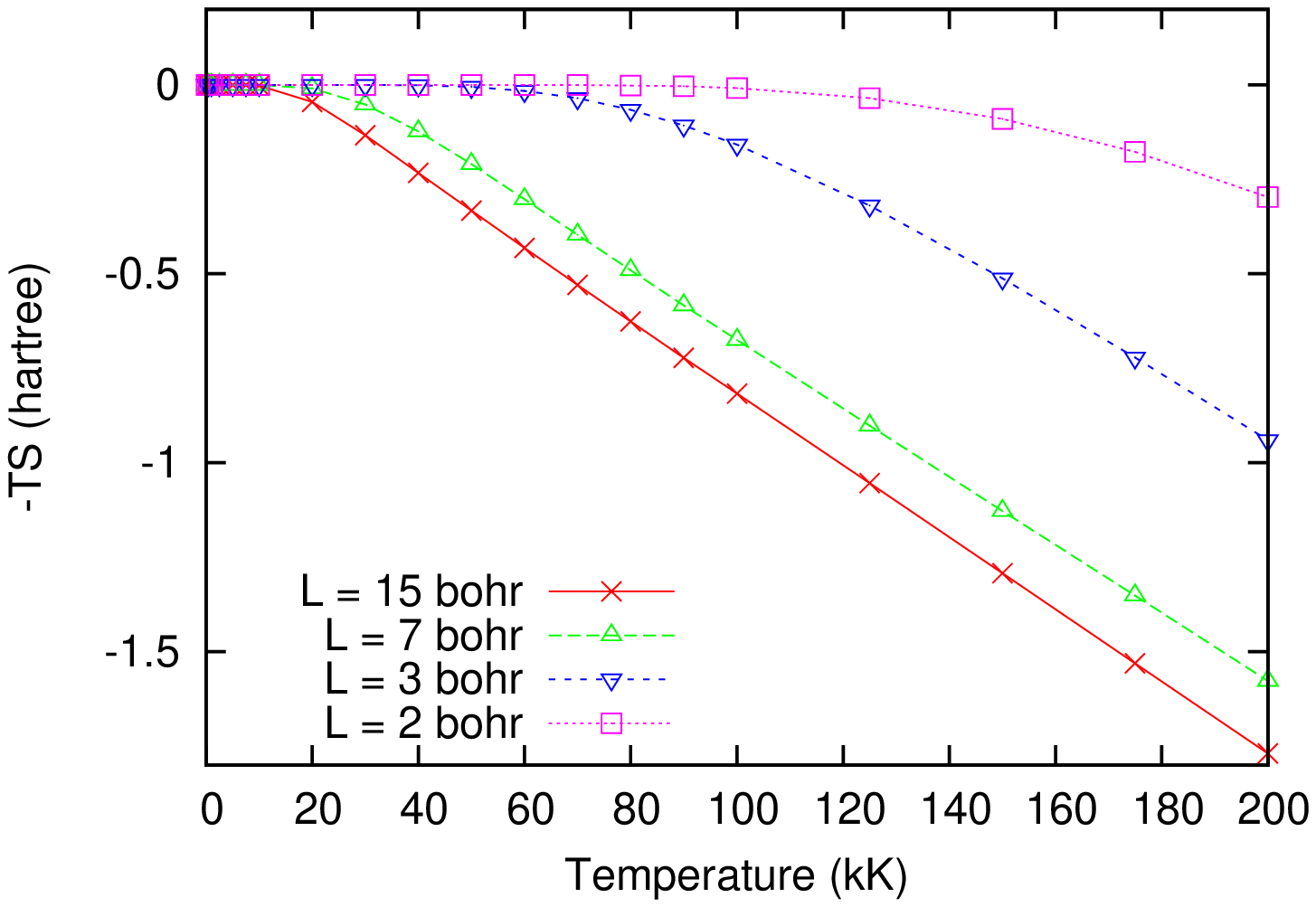}}
  \caption{Energy components of a cubically
confined single H atom vs.\ temperature.\: (a) kinetic energy,
(b) electron-nuclear energy, (c) total energy, (d) free energy,
(e) entropic energy [(d)-(c)].
Key for all is shown in (e). }
  \label{boxenergies}
\end{figure*}

Figure \ref{boxenergies}  shows the contributions to the
free energy for the single atom as a function of $\rm T$ for four cube
volumes ($L=$ 2, 3, 7, 15 bohr).
At $L = 7$ bohr, the KE is flat with
$\rm T$ at almost its $\rm T =0$ K value.  The $\rm T=50$ kK nuclear-electron attraction 
$E_{Ne}$, 
however, is much stronger for $L= 7$ bohr than for $L=15$ bohr.  By $L= 3$ bohr, 
the KE and $E_{Ne}$ are roughly equal in magnitude.

Figure \ref{boxenergies} also shows that the KE 
for the $L = 15$ bohr system falls with increasing 
temperature.  Though this might seem odd, it is  as it should 
be from virial theorem arguments for
the free atom.  The 2$s$ KE is one-fourth the 1$s$ value 
\cite{PaulingWilson}.  Finite temperature population of the
2$s$ and depopulation of the 1$s$ therefore reduces the KE with
respect to its $\rm T=0$ K value. 

Next we turn to the system of eight H atoms. We examined a symmetric
configuration in which the eight atoms were 
situated at the corners of a smaller cube, edge-length $L/2$, centered 
within the hard-wall cube, edge-length $L$.
The basis used was ten {\it s}-type GTOs centered on each atom.
Strict {\it s}-type symmetry is broken, of course,  by enforcement 
of the hard-wall 
boundary conditions. An even-tempered set of exponents  also was  
used in this case: [0.2,0.4,0.8,1.6,3.2,6.4,12.8,25.2,50.4,100.8 bohr$^{-2}$].  
As a test, the 
calculations were 
redone with  two fewer basis functions per atom; the
exponents [50.4,100.8 bohr$^{-2}$] were removed.
The two calculations agreed to 2 millihartree
 in total and component energies up to 200 kK. 
Matrix elements are calculated only once and stored, after 
which fully self-consistent 
calculations may be done at many temperatures.

\begin{figure*}
  \subfigure{\includegraphics[angle=-90,width=2.6in]{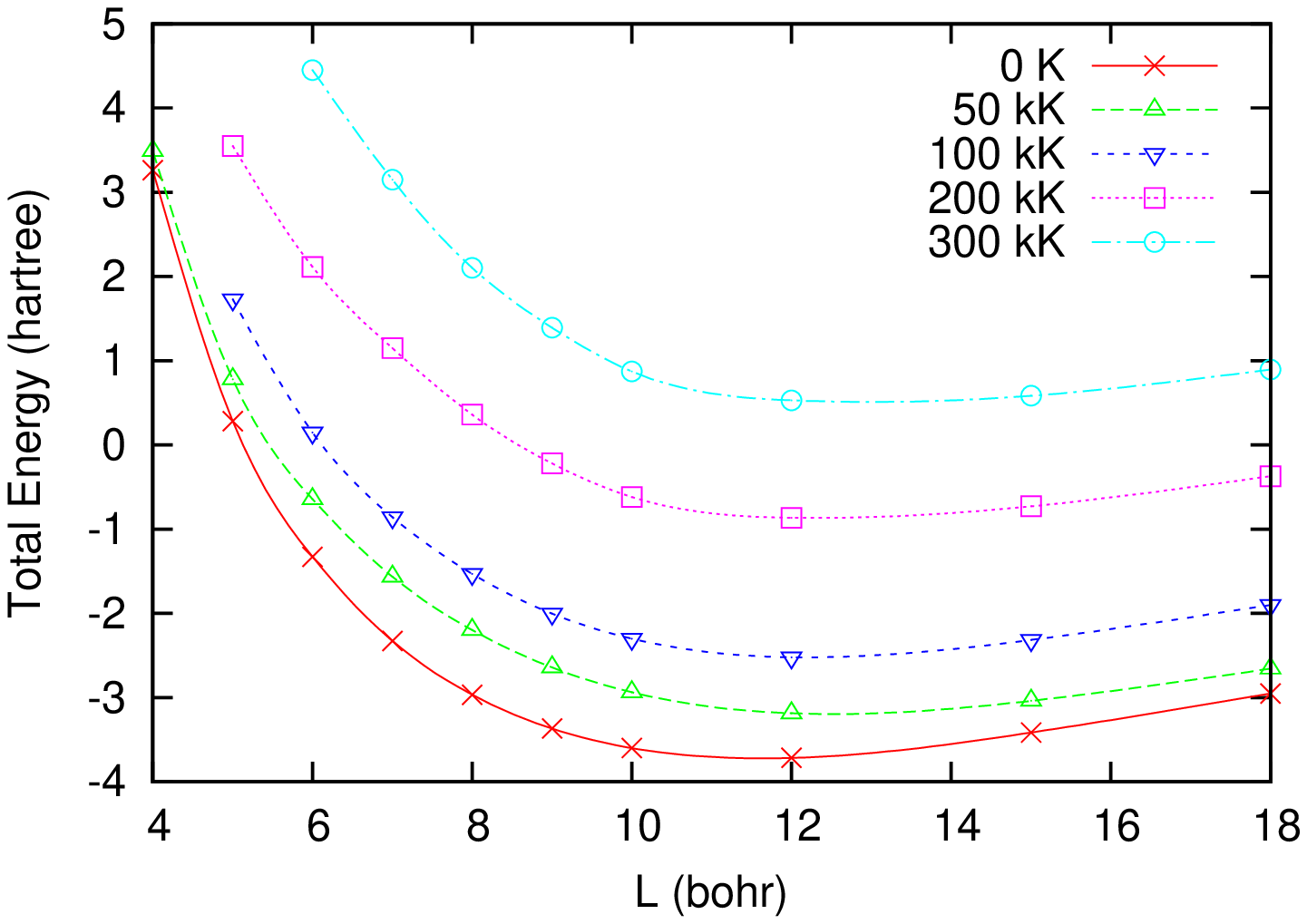}}
  \subfigure{\includegraphics[angle=-90,width=2.6in]{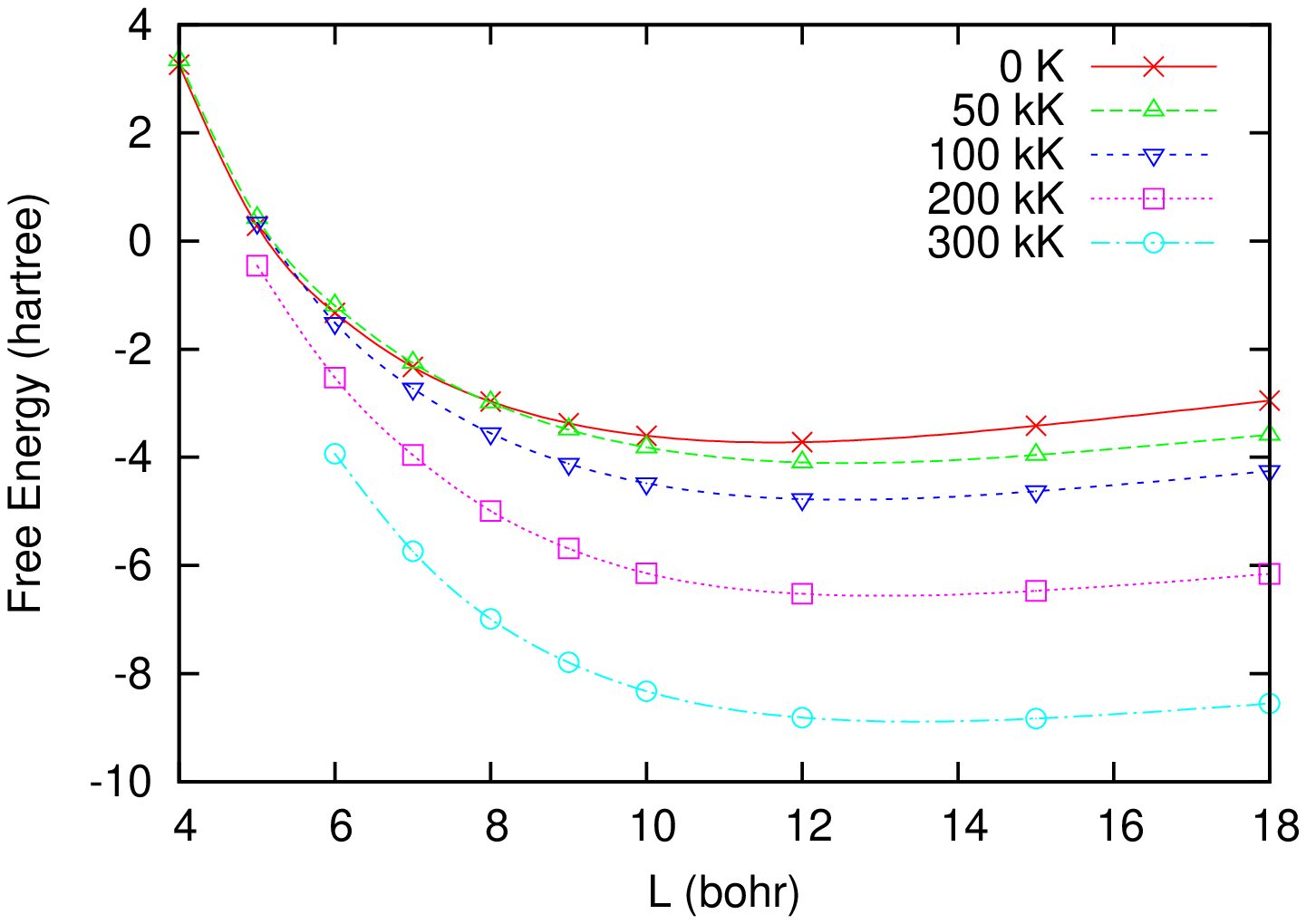}}
\caption{Zero and finite-temperature total energy and free energy
 for eight H atoms in different sized cubes:
 (a) $E={\mathcal F}_{FTHF} + {\rm T}{\mathcal S}_{FTHF}$, (b) %
$\mathcal{F}_{FTHF}$.}
\label{boxtotenergy8}
\end{figure*}

\begin{figure*}
  \subfigure[]{\includegraphics[angle=-90,width=2.6in]{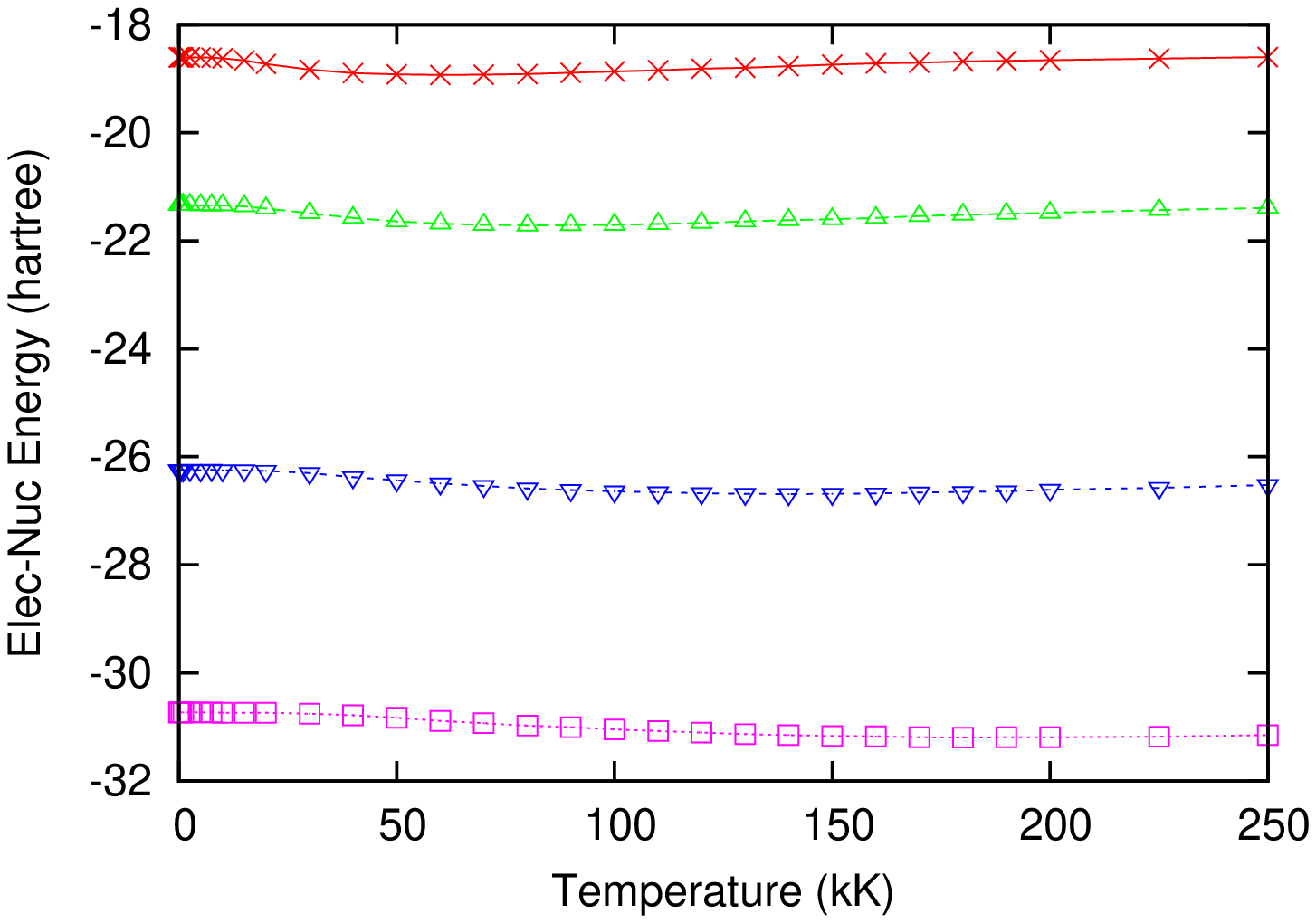}}
  \subfigure[]{\includegraphics[angle=-90,width=2.6in]{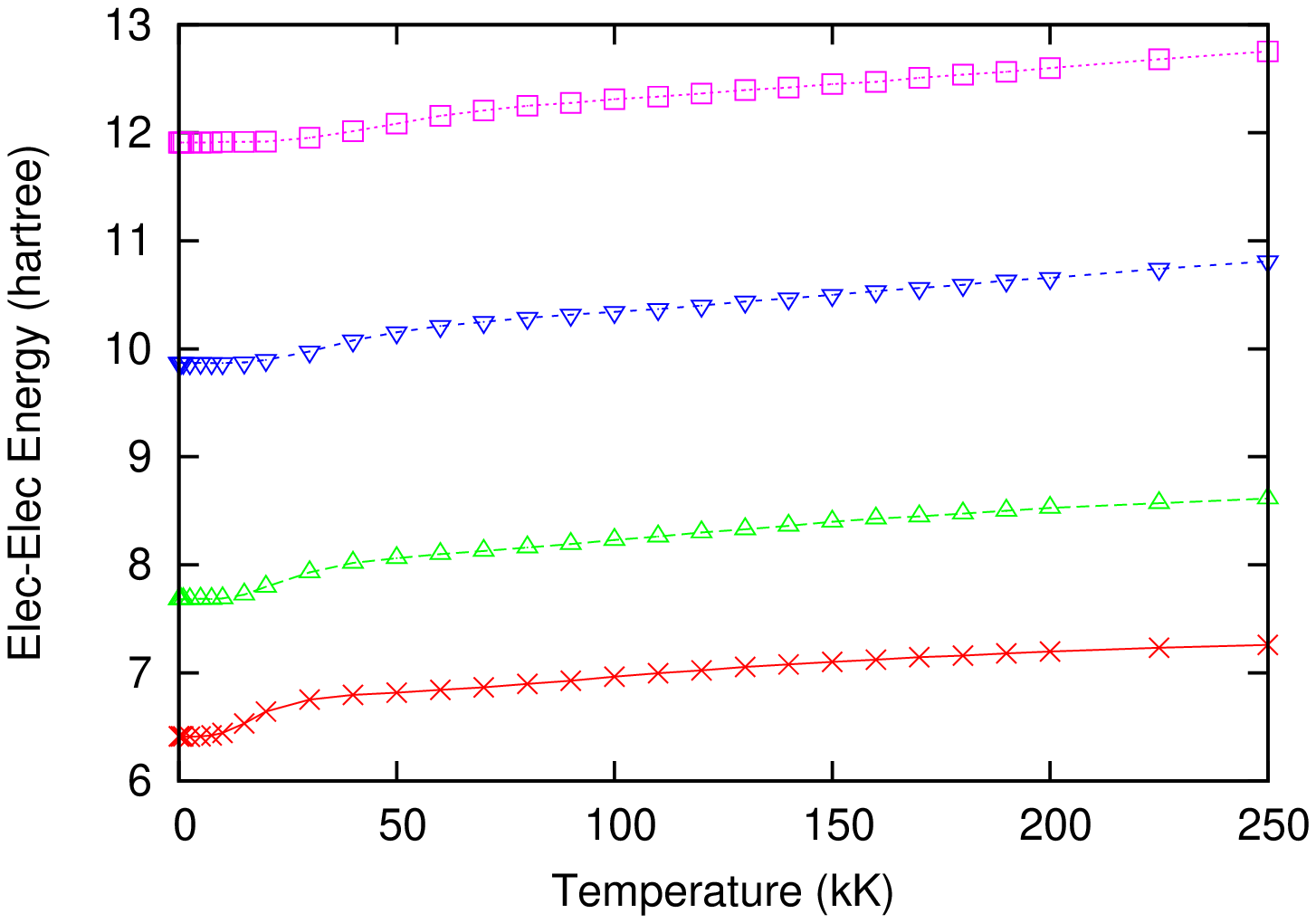}} 

  \subfigure[]{\includegraphics[angle=-90,width=2.6in]{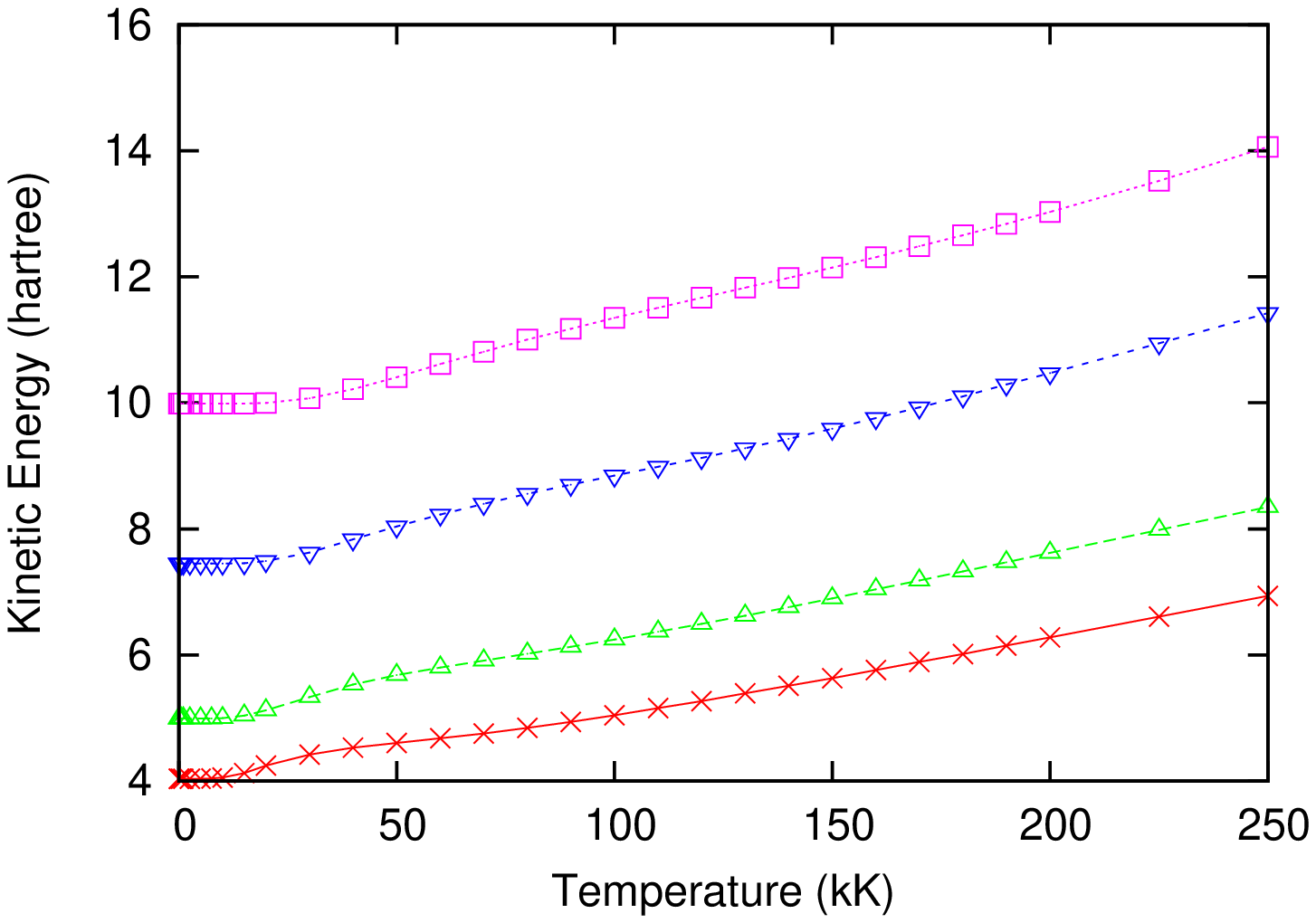}}   
  \subfigure[]{\includegraphics[angle=-90,width=2.6in]{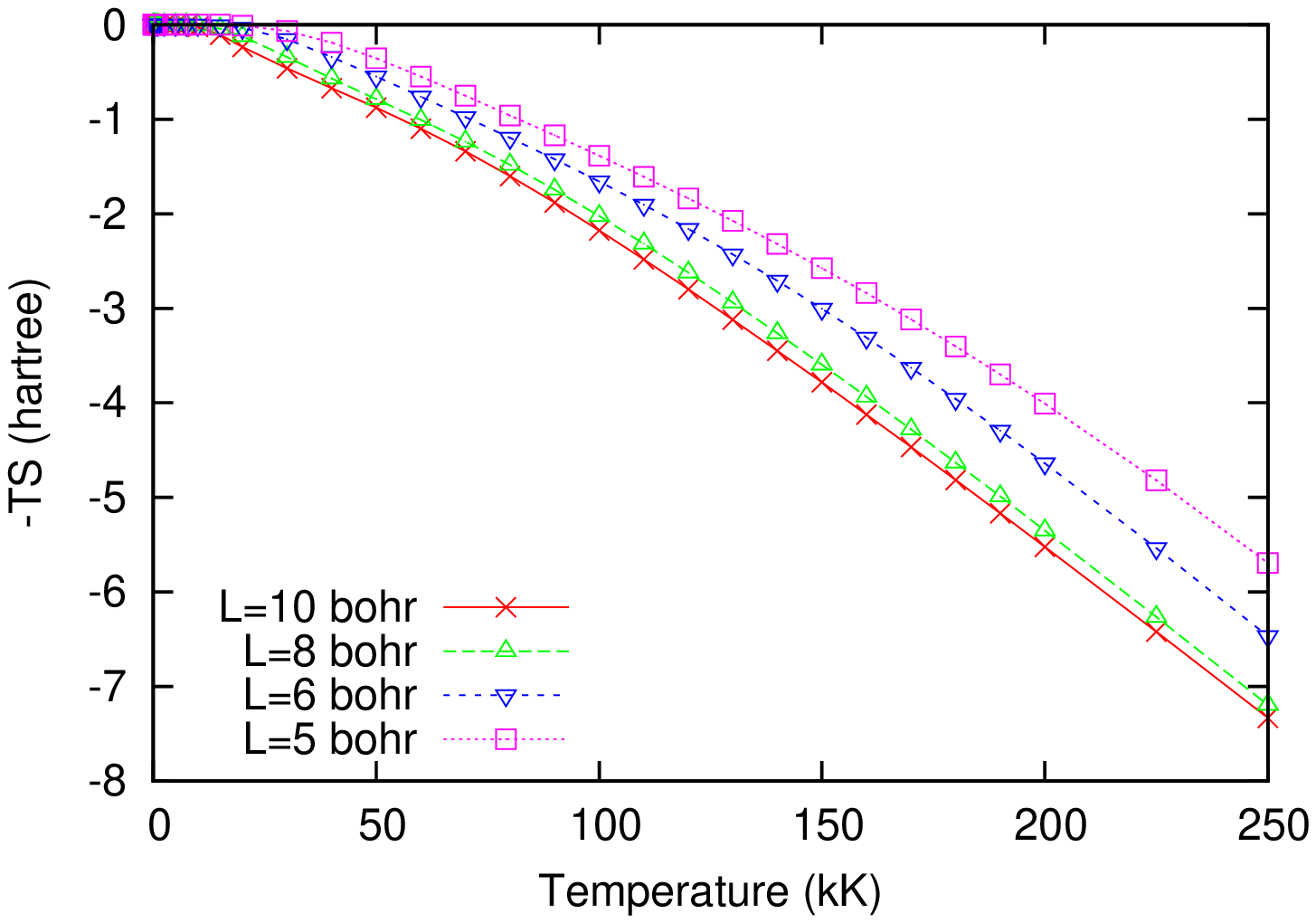}}

  \subfigure[]{\includegraphics[angle=-90,width=2.6in]{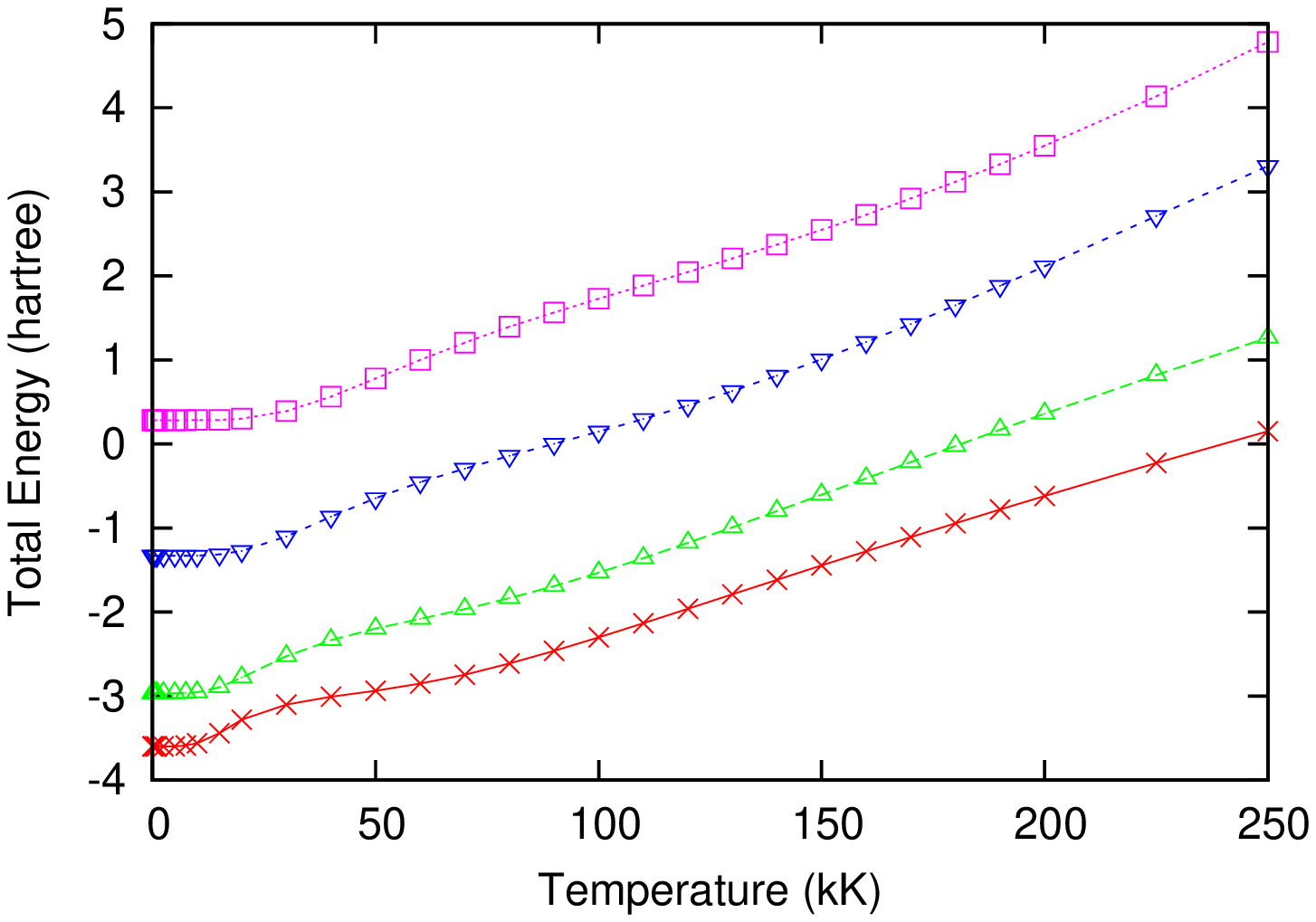}}
   \subfigure[]{\includegraphics[angle=-90,width=2.6in]{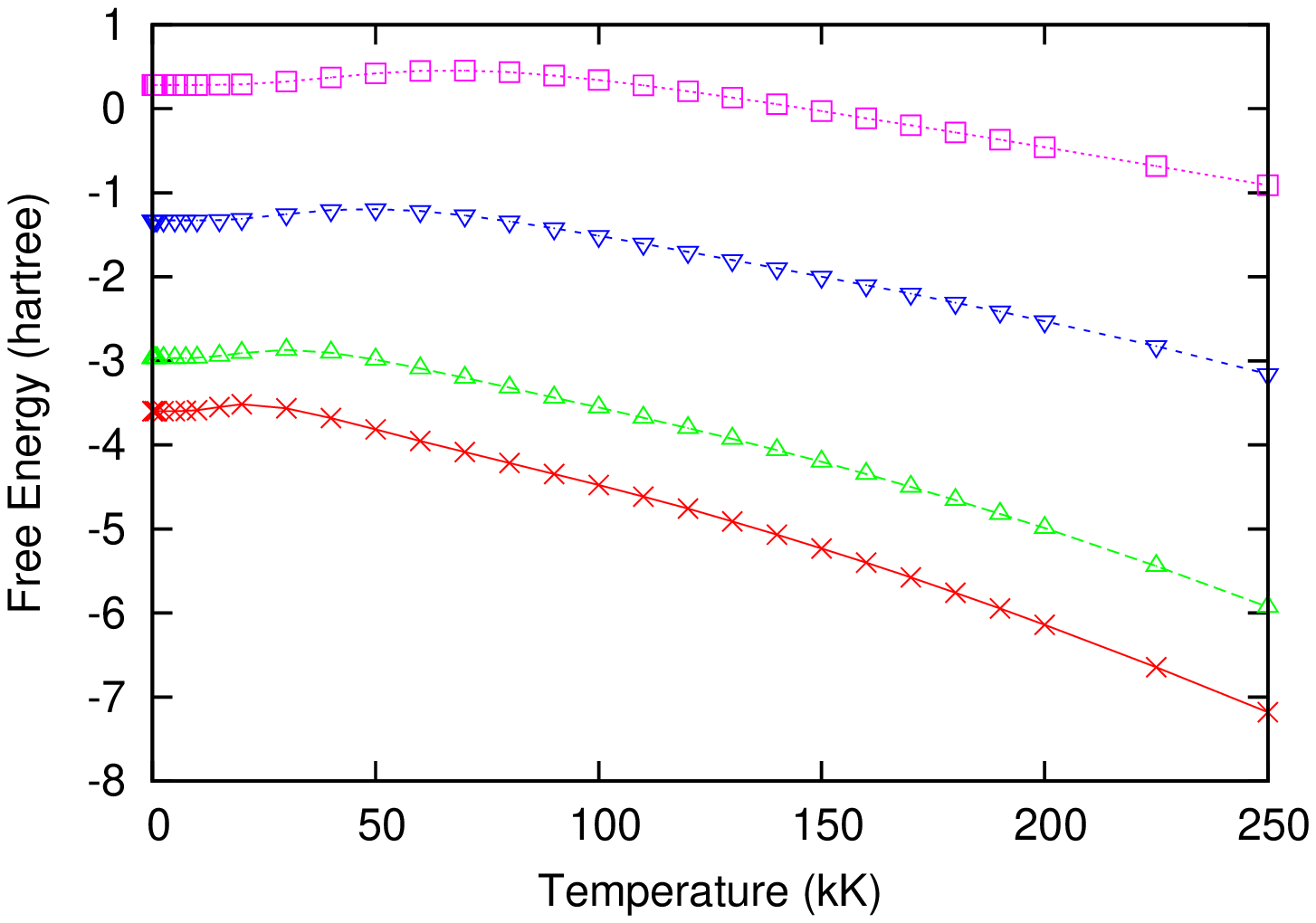}}
  \caption{Self-consistent FTHF free energy contributions for eight confined
H atoms in a cubic array as a function of $\rm T$ for four
different cube edge lengths $L$. Key for all is given in (d).}
  \label{boxenergies8-2}
\end{figure*}

Figure \ref{boxtotenergy8} shows
the total energy 
$E= {\mathcal F}_{FTHF} + {\rm T} {\mathcal S}_{FTHF}$ as a function
of $L$ for various temperatures, as well as the free energy $\mathcal{F}_{FTHF}$ itself.
The nuclear-nuclear repulsion energy is included 
(constant with respect to temperature, it varies with $L$).
Figure \ref{boxenergies8-2} shows
various components of the free energy (electron-nuclear 
Coulomb energy, electron-electron Coulomb energy including 
exchange, kinetic energy, and entropic energy). Also shown are the 
total energy and free energy, with the difference of these two 
being the entropic energy.  Again, the 
nuclear-nuclear repulsion energy is included in these two plots.
It has values of 9.118, 7.598, 5.699, 4.559 hartree for  
$L=$ 5, 6, 8, 10 bohr, respectively.  
The energies are shown as a function of 
temperature $0 \le {\mathrm T} \le 250$kK for four cube sizes, 
$L=$ 5, 6, 8, and 10 bohr.  The comparatively flat plateau 
up to roughly ${\mathrm T}=15$ kK is a 
direct consequence of Fermi-Dirac level filling. Up to about 25 kK,
the interval between the highest occupied molecular orbital (HOMO)
at zero temperature and the lowest unoccupied molecular orbital (LUMO)
is roughly constant at 0.5 hartree.  Therefore the filling ratio of those
two is roughly $\exp (-13/1.3) \approx  5 \times 10^{-5}$ or smaller 
up to about ${\mathrm T}=15$kK.  We return to this matter below.

\subsection{Comparison with approximate functionals}

Orbital-free treatment of WDM has been dominated, not surprisingly, by
local density approximations for both the KE and exchange
contributions to the free energy.  For the KE, the choice is physically
motivated by the fact that the high-pressure and/or high-temperature 
limit for a WDM system is Thomas-Fermi. This leads to 
finite-temperature Thomas-Fermi (FTTF) \cite{Latter55} ${\mathcal T}_0= \int \tau_0 \; d\mathbf{r}$
by itself, or with some fraction of von Weizs\"acker contribution (in its
zero-temperature form) ${\mathcal T}_{W}= \int \tau_W \; d\mathbf{r}$, with 
\begin{align}
  \tau_0 &=\frac{\sqrt{2}}{\pi^2\beta^{5/2}}I_{3/2}(\beta \mu)\\
  n &=\frac{\sqrt{2}}{\pi^2\beta^{3/2}}I_{1/2}(\beta \mu)\\
  \tau_W &= \frac{\left|\nabla n \right|^2}{8n}
\end{align}
where the $I$ are Fermi integrals. Note a parametrized form \cite{Perrot79} may be used to eliminate $\mu$ between $\tau_0$ and $n$.  

\begin{figure}[t]
  \includegraphics[angle=-90,width=2.6in]{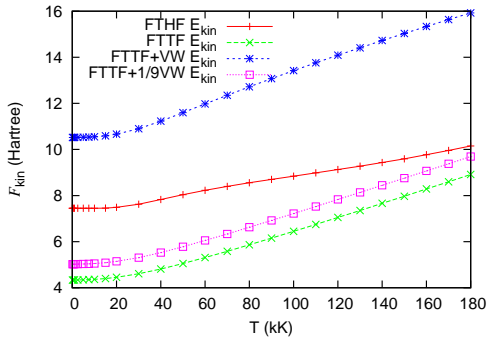}
   \caption{FTHF kinetic energy compared with finite-temperature Thomas-Fermi
KE and two forms of von Weizs\"acker augmentation of FTTF for $L =6$ bohr. See
text for details.}
  \label{KEcomparison}
\end{figure}

In Fig.\ \ref{KEcomparison} we
compare the FTHF KE as a function of $\mathrm T$ with the FTTF KE alone and
with it supplemented by both the full ${\mathcal T}_W$ and 
$(1/9){\mathcal T}_W$ for $L= 6$ bohr.  The
latter three functionals were evaluated with the FTHF density, hence are
non-self-consistent.  As is known at ${\mathrm T}=0$, pure 
Thomas-Fermi underestimates
the KE while addition of the full ${\mathcal T}_W$ overestimates it. None
of the three is close to quantitative agreement with ${\mathcal T}_{FTHF}$.
Moreover, FTTF and FTTF augmented with $(1/9){\mathcal T}_W$ can be ruled out 
from the ${\mathrm T}=0$ K comparison, since the exact 
(fully correlated) KE  must
be above  ${\mathcal T}_{FTHF}$ (from a virial theorem argument).   

\begin{figure}[t]
  \includegraphics[angle=-90,width=2.6in]{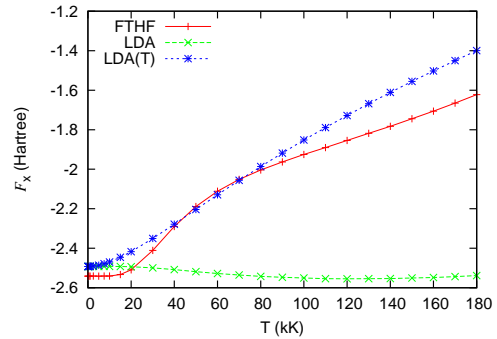}
   \caption{FTHF exchange free energy ground-state LDA and Perrot and 
Dharma-wardana temperature-dependent LDA ($L =6$ bohr). See
text for details.}
  \label{xcomparison}
\end{figure}

For the exchange contribution to the free energy, the use of ground-state 
functionals is common (recall Introduction), with the LDA being dominant.
Fig.\ \ref{xcomparison} shows the FTHF exchange contribution to the free
energy ${\mathcal F}_{x,FTHF}$ in comparison with the exchange free-energy 
generated by ground-state LDA functional 
and with the Perrot and Dharma-wardana parametrization for the 
temperature-dependent LDA functional
\cite{PerrotDw84}.  Again, this is for $L=6$ bohr and the other functionals 
are evaluated with the FTHF density.  (Note that such ``post-scf'' 
evaluation is fairly common in assessment of newly developed 
ground state functionals.  See, for example, Ref.\ 
\onlinecite{KannemannBecke09}.)   
Here one sees a marked difference: the ground-state
functional fails completely while the temperature-dependent functional has 
at least semi-quantitatively correct temperature dependence.  
%

%\subsection{Comparison with extended systems}

\begin{figure}[t]
  \includegraphics[angle=-90,width=2.6in]{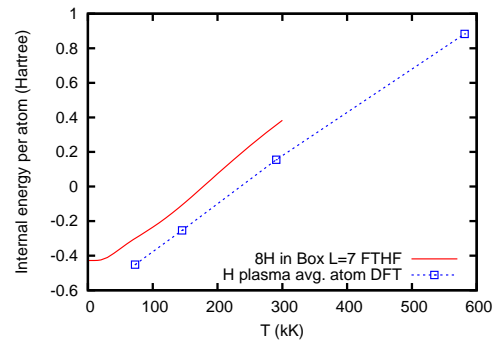}
   \caption{FTHF internal energy per atom for $L=7$ bohr, as 
compared with $r_s=2$ hydrogen plasma average atom DFT calculation.}
  \label{extcomparison}
\end{figure}

We may also 
%As this hard-walled high temperature system is unique in the
%literature, it is difficult to make comparison with previous
%works. However we may 
make some semi-quantitative comparison with a more
widely used model for extended systems at substantial T. In
Fig.\ \ref{extcomparison} we show the internal energy per atom of the
eight hydrogen atoms in the cubic symmetry arrangement at 
$L=7$ bohr (corresponds to average $r_s=2.17$), with that 
of the average atom DFT calculation
of Dharma-wardana and Perrot \cite{DWP82} at $r_s=2$. Their system
includes DFT exchange and correlation whereas we have pure 
Hartree-Fock exchange. Additional energy differnces are due to the
different boundary conditions. Decomposing the near parallel temperature
dependence into those components, however, is a task  outside the
scope of the present work.   We can already see one aspect.  
Though the kinetic energy is the
major contributor to the change in total energy as a function of temperature
(recall Fig.\ \ref{boxenergies8-2}), the change due to
electron-electron interaction, including exchange, is at least a third
that of the kinetic energy.

\section{Discussion and Conclusions}

Comparison, evaluation, and betterment of functionals for WDM simulations
is the long-term motive of this work.  As just shown, even at this initial state
of development (Hartree-Fock, small particle number, no molecular dynamics),
the approach gives insight regarding that task.  There are some specific
issues worth discussion also.

\subsection{One-particle spectrum effects}

It is well known that zero-temperature HF calculations over-estimate
both band
gaps in solids and the so-called HOMO-LUMO (highest occupied and lowest
empty molecular orbital respectively) gaps of finite systems. 
This happens because the 
occupied HF orbitals are free of self-interaction, but the unoccupied 
ones are not. As temperature increases in the 
FTHF scheme, however, levels unoccupied at 
T = 0 K become increasingly occupied and shifted. 
In the self-consistent solution of the HF problem in a basis, 
those levels then contribute to the Hamiltonian matrix (``Fock matrix'' 
in quantum chemistry terminology).  Thus there are two
questions to address.  What is the extent to
which FTHF exhibits HOMO-LUMO gap over-estimation similar to that of 
ground-state HF?  At what  fractional occupancy is an energy level 
changed from an
overestimated virtual to a more properly estimated partially occupied one? 

\begin{figure}
    \includegraphics[angle=-90,width=2.6in]{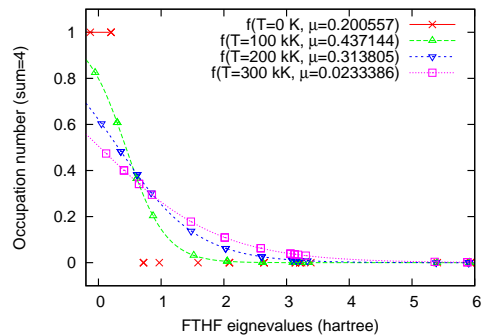}
    \caption{Fermi distribution for eight
 H atoms in a box of size $L=6$ bohr. Points may represent more than
 one state due to energy degeneracies. }
    \label{fermi-dist}
  \end{figure}

\begin{figure}
    \includegraphics[angle=-90,width=2.6in]{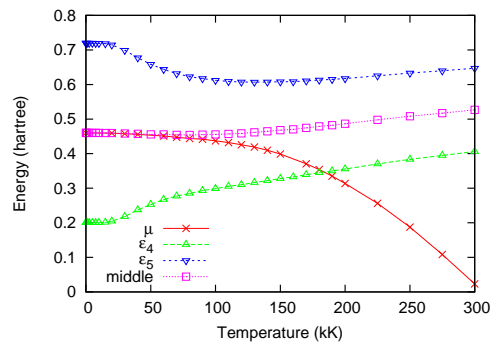}
    \caption{Chemical potential $\mu$ and the fourth and fifth energy levels.
The curve labeled ``middle'' is half way between $\varepsilon_4$ and $\varepsilon_5$. 
    \label{mu2} }
  \end{figure}

\begin{figure}
    \includegraphics[angle=-90,width=2.6in]{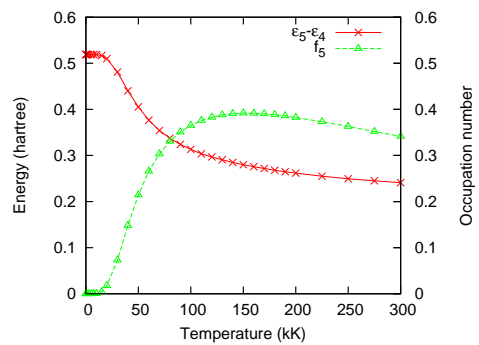}
    \caption{Energy difference between the fourth and fifth energy levels along with the occupation number of the fifth energy level.
    \label{diff} }
  \end{figure}

For specificity, we consider the eigenspectrum of a single, moderately
compressed eight-atom cube with $L=6$ bohr.
Figure \ref{fermi-dist} shows the Fermi distribution of the single-particle 
energies for four temperatures. Note that 
these distributions are for one spin.  Also keep in mind that the
cubical symmetry causes the lowest four one-electron orbitals to 
group as singly degenerate and triply degenerate (a1g, t1u in crystal
field notation).  At zero temperature therefore, only two
points are shown with unit occupancy, but the higher energy point
corresponds to three degenerate HOMO states (indexed as 2, 3, 4).  
The LUMO is the degenerate states 5, 6, and 7, with the singly 
degenerate state 8 above them (again as would be expected from 
a cubical crystal field).  For simplicity of discussion, 
the HOMO and LUMO (at ${\rm T} =0$) are labeled  $\varepsilon_4$ and
$\varepsilon_5$ respectively. One can see that the spacing 
between $\varepsilon_4$ and $\varepsilon_5$ decreases with
temperature. This difference is shown directly in Fig.\ \ref{diff}
along with the occupation number for the $\varepsilon_5$ level.
The continuous curves in Fig.\ \ref{fermi-dist} show the
Fermi function with calculated chemical potential $\mu$.  The discrete 
points mark the input energies to the calculation of $\mu$. Then 
in Fig.\ \ref{mu2}, $\mu$ is plotted as a function of $\rm T$.  Observe
that the chemical potential is nearly mid-way 
between  $\varepsilon_4$ and
$\varepsilon_5$  up to just below $50$ kK.   This behavior is exact at zero
temperature.  

Those total energy and eigenspectrum results together resolve 
the matter of the behavior of what would be virtual states at zero
temperature in FTHF. First, examination
of the kinetic and total energy plots for all box sizes shows that there 
is a change
in the form of the temperature dependence at roughly $20$ kK. That 
change is
complemented by the change in the spacing between the $\varepsilon_4$ and
$\varepsilon_5$ levels.  They are essentially static for lower temperatures, 
then begin to change abruptly well below $50$ kK, and then change 
more moderately at 
higher temperatures. Thus, above $50$ kK, states corresponding to 
zero-temperature virtuals are 
sufficiently incorporated in the interaction terms to make a material
modification of ${\rm T} = 0$ behavior.
However, as the temperature is decreased below roughly $50$ kK, 
the FTHF Coulomb and exchange terms increasingly are dominated 
by the ${\rm T} = 0$
occupied levels, which therefore keeps the lightly occupied higher 
energy levels artificially high. This is not a basis issue, but an issue with
discrete eigenstates. In a solid such as jellium or a metal with a
continuum of states, this should not be an issue, but for a system with
energy gaps the issue remains. 


\subsection{Other findings and considerations}

The preceding discussion about depopulation and repopulation of levels
relative to the ground-state HF HOMO-LUMO gap illustrates a broader
challenge for the construction of approximate functionals for the
various contributions to the free energy.  Whether explicit or
implicit, such approximate functionals correspond to restricting the
required traces to specific classes (or sub-sets of classes) of state
functions.  The consequence of such a restriction is to incorportate
the spectral properties of that class into the approximate
functionals.  For example, in FTHF that class is single Slater
determinants constructed with respect to ground-state HF
minimization. Although we have not attempted its construction here, in
principle there is an FTHF free-energy functional.  It would have
exactly the same problem with a plateau in its T-dependence as we have
found here.
  
The small number of particles is another issue.  For sufficiently 
large numbers, all standard ensembles (grand canonical, canonical,
micro-canonical) give the same thermodynamics.  Fluctuations
characteristic of small particle counts can degrade that relationship.
Such issues are endemic to any computational study, especially when 
a large temperature and pressure domain such as characterizes WDM is 
involved.  In that sense, the present study is not much worse off
than most modern WDM simulations  with small numbers of electrons 
\cite{HostFrenchRedmer11}.   At the least, we have an even-handed
comparison of different methods ({\it e.g.}, the comparison of
functionals given above) for a given number of electrons and of ions.
The main issue regarding particle count is computational cost.

Clearly we have shown that
the tGTO basis is feasible and effective.  As is typical of GTO basis
methods, the computational cost  is essentially entirely in the 
cpu time for the calculation of the two-electron integrals. The eight-atom 
systems described are calculated with 64 or 80 total basis functions, making
diagonalization trivial. A simple double array of all $N^4$
two-electron integrals only occupies 128 or 312.5 MB respectively,
storable in memory. In practice we calculated $N^4/ \lambda$ integrals, with
$\lambda=$7.60, 7.68 and not 8 due to the looping procedure we used.  While 
for  the cubic system discussed here, this calculation requirement could 
have been reduced further by exploiting 
symmetry, we need the capability to explore other, lower symmetry 
geometries. Note however that
if the box size or the atomic positions are changed, all integrals
affected by the change must be recalculated.

On a modern desktop processor (Intel Core i5 650 at 3.2 Ghz) one
two-electron integral can be calculated in about 29.1 ms. So the times to
calculate all integrals for a 64- or 80-orbital basis would be 17.84
hr and 43.13 hr. The integrals are calculated independently, so can be
parallelized effectively. Calculations reported in this work were done on
the University of Florida High Performance Computing Center Linux clusters.

Though these are quite acceptable costs for fixed
geometries and small numbers of ions and electrons, the 
burden becomes formidable for direct application in Born-Oppenheimer 
molecular dynamics. We are currently working on ways to ameliorate that
problem.

\section{Acknowledgement}

\label{sec6} This work was supported under US DOE Grant DE-SC0002139.
We acknowledge useful conversations with Mike Murillo early in this
work and continuing discussions with our colleagues Jim Dufty, Keith
Runge, and Valentin Karasiev.  We also thank the University of Florida
High Performance Computing Center for computational support.  

\appendix

\section{Correction for Piecewise Integrals for the Basis Functions}

{F}rom the definition of the basis functions in Eqs.\ \eqref{xstype} and
\eqref{sgtotruncated}, it follows that derivatives of the basis function
may not be continuous at $x_c$, the center of the function.  A 
discontinuity of  the second
derivative would, of course, be significant for the kinetic energy. 
The issue is whether the kinetic energy matrix elements can be 
evaluated piecewise, as is the case with the
overlap, nuclear-electron, and electron-electron integrals.  We may examine 
this issue by writing the basis formally with Heaviside functions, as
follows:
\begin{equation}
  \varphi =  \left[ \theta (x) - \theta(x-x_1) \right] \varphi_0 +
  \left[ \theta (x-x_1) - \theta(x-L) \right] \varphi_L  
\label{A1}
\end{equation}
For the tGTO basis, the identification from Eq.\ \ref{sgtotruncated} is 
\begin{align}
  &\varphi_0 = a_0 \left( g^n(x) - \delta_0 \right) \nonumber \\
  &\varphi_L = a_L \left( g^n(x) - \delta_L \right) \nonumber \\
  &g^n(x) = (x-x_1)^n e^{-\alpha(x-x_1)^2}  \;,\quad x_1 = x_c  \; 
\label{varphidefn}
\end{align}
The derivatives are 
\begin{align}
  \frac{\partial \varphi}{\partial x} 
  =& \left[ \delta (x) - \delta(x-x_1) \right] \varphi_0 + %
\left[ \theta (x) - \theta(x-x_1) \right] \varphi_0^{\prime}  \nonumber \\
&+ \left[ \delta (x-x_1) - \delta(x-L) \right] \varphi_L \nonumber\\
&+\left[ \theta (x-x_1) - \theta(x-L) \right] \varphi_L^{\prime} 
\label{firstderiv}
\end{align}
and
\begin{align}
  \frac{\partial^2 \varphi}{\partial x^2} 
  =& \left[ \delta^{\prime} (x) - \delta^{\prime}(x-x_1) \right] \varphi_0 %
 + 2 \left[ \delta (x) - \delta(x-x_1) \right] \varphi_0^{\prime} \nonumber \\
&+ \left[ \theta (x) - \theta(x-x_1) \right] \varphi_0^{\prime \prime} %
  \nonumber \\
&+ \left[ \delta^{\prime} (x-x_1) - \delta^{\prime}(x-L) \right] \varphi_L \nonumber \\%
&+ 2 \left[ \delta (x-x_1) - \delta(x-L) \right] \varphi_L^{\prime} \nonumber \\
 &+ \left[ \theta (x-x_1) - \theta(x-L) \right] \varphi_L^{\prime \prime} \; 
\label{2ndderiv}
\end{align}

Now consider a generic kinetic energy matrix element involving the 
foregoing 
function and another, similar basis function $\chi$ with
left and right constituents $\chi_0$, $\chi_L$, centered at $x_2$. 
Without loss of generality, take $x_1 < x_2$. Then
\begin{equation}
  \int_0^L \chi  \frac{\partial^2 \varphi}{\partial x^2} \; dx := I_A + I_B
\label{basicKEint}
\end{equation}
The terms of the second derivative with the Heaviside functions
contribute just the piecewise integration $I_A$, while the delta
function and first derivative delta function terms contribute
$I_B$. Of the terms in Eq.\ \ref{2ndderiv} only those at $x_1$ 
contribute, as the constituents $\chi_0$, $\chi_L$, $\varphi_0$, 
and $\varphi_L$ go to zero at $x=0$ and $x=L$.  Thus 
\begin{align}
I_B =&  \int \chi_0 \left[ \delta^{\prime}(x-x_1) %
 \left(\varphi_L -\varphi_0\right) \right. \nonumber \\%
&\left. + 2 \delta(x-x_1) \left( \varphi_L^{\prime}-\varphi_0^{\prime} \right) \right] %
 \; dx  \; .
\label{IB}
\end{align}
{F}rom the definition of $\delta^{\prime}$ this expression becomes
\begin{align}
I_B =&  \int \chi_0 \left[-\delta(x-x_1) %
\left(\varphi_L^{\prime} -\varphi_0^{\prime}\right) \right. \nonumber \\%
&\left. +2 \delta(x-x_1) \left( \varphi_L^{\prime}-\varphi_0^{\prime} \right) \right] %
 - \chi_0^{\prime} \delta(x-x_1) \left(\varphi_L -\varphi_0\right)\; dx \;\;
\label{IBintermed}
\end{align}
which reduces to
\begin{equation}
  I_B = \left[ \chi_0 \left(\varphi_L^{\prime} -\varphi_0^{\prime}\right) %
- \chi_0^{\prime} \left(\varphi_L -\varphi_0\right) \right] \bigg|_{x=x_1} \; 
\label{IBend}
\end{equation}

For the case of $x_1 = x_2$, that is, for diagonal terms or functions
that have the same center,  $\chi_0^{\prime}$ must be replaced in
Eq.\ (\ref{IBintermed}) by the analog of 
Eq.\ (\ref{firstderiv}), with the result
\begin{widetext}
\begin{align}
  I_B = \chi_0 \left(\varphi_L^{\prime} -\varphi_0^{\prime}\right)%
 \bigg|_{x=x_1}- \int &\delta(x-x_1) \left(\varphi_L -\varphi_0\right)%
 \left[ \left[ \delta (x) - \delta(x-x_1) \right] \chi_0 %
+ \left[ \theta (x) - \theta(x-x_1) \right] \chi_0^{\prime}  \right. %
\nonumber \\
&+ \left. \left[ \delta (x-x_1) - \delta(x-L) \right] \chi_L %
+ \left[ \theta (x-x_1) - \theta(x-L) \right] \chi_L^{\prime} \right] \; dx %
\; 
\label{I1intermed2}
\end{align}
\end{widetext}
Note $\chi_0$ in the first term follows because the functions themselves 
are continuous,while the first and second derivatives may not be. For the same reason, it follows that the remaining integral in Eq.\ (\ref{I1intermed2})  and the second term of Eq.\ (\ref{IBend}) vanish. Thus so long as the functions are continuous, the correction to
the piecewise kinetic energy integral is simply
\begin{equation}
  I_B = \chi (\varphi_L^{\prime}-\varphi_0^{\prime})\bigg|_{x=x_1}
\label{IBke}
\end{equation}
and so long as the first derivative is continuous, this reduces to zero.

With continuity of the basis functions enforced by construction,
only the first derivative needs to be examined for a possible correction
to simple 
piecewise integration. For the basis defined in
Eqs.\ (\ref{sgtotruncated}), those corrections are
\begin{align}
  g^0(x)&=e^{-\alpha(x-x_1)^2} &\quad&   I_B = 0 \nonumber \\
  g^1(x)&=(x-x_1) e^{-\alpha(x-x_1)^2} &\quad&   I_B = \chi (a_L-a_0) \nonumber \\
  g^2(x)&=(x-x_1)^2 e^{-\alpha(x-x_1)^2} &\quad&  I_B = 0  \; 
\label{discontContrib}
\end{align}
Basis functions $g^n(x)$ with higher powers of the prefactor $(x-x_1)$ all 
have continuous first and second derivatives at $x_1$. In fact, those 
derivatives are all zero. So only the $p$-type basis functions, $(n=1)$, 
have a kinetic energy matrix element contribution beyond 
that given by piecewise integration.

\section{Finite-Range Gaussian Integrals}
%%% SBT inserted Travis' revised Appendix B (28 June 2011) here.
Following the methods of Boys 
\cite{Boys50,Boys60,Singer60,LongstaffSinger60,Gill94}, 
we use the transform of the Coulomb potential to separate the Coulomb integrals 
into one-dimensional Cartesian pieces:
\begin{align}
  &\frac{1}{\left|{\bf r} - {\bf R_N}\right|}=\frac{1}{\sqrt{\pi}} \int_{-\infty}^{\infty} e^{-s^2 ({\bf r}-{\bf R_N})^2} ds \nonumber \\
&=\frac{1}{\sqrt{\pi}} \int_{-\infty}^{\infty} e^{-s^2 (x- X_N)^2}e^{-s^2 (y-Y_N)^2}e^{-s^2 (z-Z_N)^2} ds \; .
\label{FTCoulomb}
\end{align}
Hence all Coulomb integrals require integration over the transform 
variable $s$, which is done by Gauss-Laguerre quadrature.

For 1D primitive tGTOs, we note the required finite range $(a,b)$
integrals are of the form 

\begin{equation}
I_n=\int_a^b x^n e^{-\alpha(x-x_c)^2} \; dx
\end{equation}
This simply transforms to 
\begin{equation}
  I_n=\int_{a-x_c}^{b-x_c} (x^{\prime} + x_c)^n e^{-\alpha x^{\prime 2}} \; dx^{\prime}
\end{equation}
This result leaves us needing to compute 
\begin{equation}
  J_n=\int_{a^\prime}^{b^{\prime}} x^n e^{-\alpha x^2} \; dx
\end{equation}
Integrating by parts we find 
\begin{equation}
  J_n=-x^{n-1}\frac{e^{-\alpha x^2}}{2 \alpha}\bigg|_{a^\prime}^{b^{\prime}} + \frac{n-1}{2\alpha}J_{n-2}
\end{equation}
So with the initial two integrals, we may find the higher-order 
integrals by recursion:
\begin{align}
J_0 & = \int_{a^\prime}^{b^{\prime}} e^{-\alpha x^2} \; dx =  
 \frac{\sqrt{\pi}}{2\sqrt{\alpha}} \erf \left(\sqrt{\alpha} x \right)\bigg|_{a^\prime}^{b^{\prime}} \\
J_1 & = \int_{a^\prime}^{b^{\prime}} x e^{-\alpha x^2} \; dx =
-\frac{e^{-\alpha x^{2}}}{2 \alpha}\bigg|_{a^\prime}^{b^{\prime}}
\label{Jinitial}
\end{align}

All two-center (overlap, nuclear-electron, kinetic)
integrals reduce to expressions in terms of $I_n$. 
After two applications of  the Gaussian product rule, four-center 
(two-electron) integrals reduce to terms of the form
\begin{equation}
  \int_{a_2}^{b_2} \int_{a_1}^{b_1} x_1^n x_2^m e^{-\alpha_1(x_1-x_i)^2} e^{-\alpha_2(x_2-x_j)^2} 
e^{-s^2(x_1-x_2)^2} \; dx_1 dx_2
\label{intermedC}
\end{equation}
Two further applications of the product rule bring us to the form 
\begin{equation}
  \int_{a_2}^{b_2} \int_{a_1}^{b_1} x_1^n x_2^m e^{-\kappa_1(x_1-R_1(x_2))^2} e^{-\kappa_2(x_2-R_2)^2}\; dx_1 dx_2
\label{intermedD}
\end{equation}
Here the integral over $x_1$ may be evaluated as $I_n$, so we are left with 
\begin{equation}
  \int_{a_2}^{b_2} I_n(x_2)\bigg|_{x_1=a_1}^{x_1=b_1} x_2^m e^{-\kappa_2(x_2-R_2)^2} \; dx_2 
\label{intErf}
\end{equation}
which we evaluate by Gauss-Legendre quadrature.

%%%%%%%%%%%%%%%%%%%%%%%%%%%%%%%%%%%%%%%%%%%%%%%%%%%%%%%%%%%%

\end{document}